\documentclass[twocolumn,floatfix,prb,aps,showpacs]{revtex4-1}
\usepackage{graphicx,amsmath,amssymb,color}
\usepackage{nicefrac}
\usepackage[titletoc,title]{appendix}

\newcommand{\be}{\begin{equation}}
\newcommand{\ee}{\end{equation}}

\newcommand{\ba}{\begin{eqnarray}}
\newcommand{\ea}{\end{eqnarray}}

\begin{document}

\title{Parton construction of a wave function in the anti-Pfaffian phase}
\author{Ajit C. Balram$^{1}$, Maissam Barkeshli$^{2}$, and Mark S. Rudner$^{1}$}
\affiliation{$^{1}$Niels Bohr International Academy and the Center for Quantum Devices, Niels Bohr Institute, University of Copenhagen, 2100 Copenhagen, Denmark}
\affiliation{$^{2}$Condensed Matter Theory Center and Joint Quantum Institute, Department of Physics, University of Maryland, College Park, Maryland 20472 USA}
\date{\today}

\begin{abstract} 
In this work we propose a parton state as a candidate state to describe the fractional quantum Hall effect in the half-filled second Landau level. The wave function for this parton state is $\mathcal{P}_{\rm LLL} \Phi_{1}^3[\Phi_{2}^{*}]^{2}\sim\Psi^{2}_{2/3}/\Phi_{1}$ and in the spherical geometry it occurs at the same flux as the anti-Pfaffian state. This state has a good overlap with the anti-Pfaffian state and with the ground state obtained by exact diagonalization, using the second Landau level Coulomb interaction pseudopotentials for an ordinary semiconductor such as GaAs. By calculating the entanglement spectrum we show that this state lies in the same phase as the anti-Pfaffian state. A major advantage of this parton state is that its wave function can be evaluated for large systems, which makes it amenable to variational calculations. In the appendix of this work we have numerically assessed the validity of another candidate state at filling factor $\nu=5/2$, namely the particle-hole-symmetric Pfaffian (PH-Pfaffian) state. We find that the proposed candidate wave function for the PH-Pfaffian state is particle-hole symmetric to a high degree but it does not appear to arise as the ground state of any simple Hamiltonian with two-body interactions.
\end{abstract}
\maketitle

\section{Introduction}

Two dimensional electronic systems can exhibit a wide variety of interesting and exotic quantum phases of matter. In particular, when a strong magnetic field is applied, electron-electron interactions give rise to ``fractionalized'' phases in which the fundamental excitations over the ground state are characterized by quantum numbers that differ from those of the constituent electrons. Notably, quasiparticles and quasiholes in fractional quantum Hall (FQH) systems have been shown to carry fractional values of the electron charge~\cite{Laughlin83, de-Picciotto97, Dolev08, Venkatachalam11}, and are expected to obey ``anyonic'' exchange or braiding statistics~\cite{Wilczek82, Arovas84, Wen91, Read00}. The excitations at certain filling factors such as $\nu = 5/2$~\cite{Willett87,Moore91, Tiemann12} and $\nu = 12/5$~\cite{Read99, Xia04, Rezayi09, Kumar10, Pakrouski16} have even been predicted to feature {\it non-Abelian} braiding statistics; such states have been proposed as potential platforms for implementing topologically-protected fault-tolerant quantum computation~\cite{Kitaev03, Nayak08}, stimulating great interest in their theoretical characterization and experimental realization.

From a theoretical point of view, the study of FQH phases presents a number of challenges. While the problem can be considerably simplified by projecting the single particle Hilbert space to the lowest Landau level (LLL), the massive many-body degeneracy at fractional filling renders the problem essentially non-perturbative. Numerical ``exact diagonalization'' is useful, but even in the restricted space of the LLL- can only be carried out for rather small systems. Considerable insight can often be obtained through the construction of variational ``trial'' wave functions. Prominent early examples include the Laughlin~\cite{Laughlin83} and Halperin~\cite{Halperin84} states, which incorporate electron-electron correlations via simple pair-wise Jastrow factors in their respective many-body wave functions. The composite fermion paradigm~\cite{Jain89} provides a remarkably successful systematic framework for constructing correlated ground and excited state trial wave functions at fractional filling, built from integer quantum Hall (IQH) states of electrons bound to an even number of vortices. The parton construction first introduced by Jain~\cite{Jain89b}, which we focus on below, generalizes the composite fermion approach to describe even more exotic states.

Importantly, whole new classes of states, at different filling fractions from those found by the methods above, can be obtained through the operation of particle-hole conjugation: by taking a LLL many-body wave function at filling fraction $\nu$, a state at filling fraction $1 - \nu$ (or $2 - \nu$ for non-spin-polarized states) is generated by replacing filled states with empty ones, and vice-versa, in the LLL Fock-space representation of the state. In the limit of large magnetic field this is a relevant operation since with extremely high probability only the LLL states are occupied. Even when a given variational ansatz is easy to evaluate, it is generically hard to construct the state related to it by particle-hole conjugation. This is so because the operation of particle-hole conjugation requires knowing the exact expansion of the state in the full Hilbert space. Thus we are motivated to seek simple-to-evaluate variational wave functions for such ``particle-hole conjugate states,'' to facilitate their further investigation. 

In this paper we focus primarily on filling factor $\nu = 5/2$. We propose a parton wave function at this filling factor, and show that it lies in the same phase as the so-called anti-Pfaffian state~\cite{Levin07,Lee07}, which is the particle-hole conjugate of the Pfaffian state~\cite{Moore91}. Our parton state has a good overlap with both the anti-Pfaffian state and with the numerically-exact ground state obtained using the Coulomb interaction pseudopotentials for the second LL of ordinary semiconductors such as GaAs. Unlike the anti-Pfaffian state, the parton state can be easily evaluated for very large systems, thus making it amenable to variational calculations. 

Below, in Sec.~\ref{sec:parton}, we first give some background on parton states and then describe the ansatz of our candidate parton state at $\nu = 5/2$. Then in Sec.~\ref{sec:candidate_parton} we describe some of the properties of this state; in particular we show that it lies in the anti-Pfaffian phase. In Sec.~\ref{sec:applicatons_extensions} we demonstrate the utility of our construction by calculating the pair-correlation function and static structure factor of this state for a large number of particles. We further extend the parton construction to a larger family of states. In Sec.~\ref{sec:discussion} we discuss our results and provide an outlook for the future.

\section{Parton ansatz}\label{sec:parton}
We now briefly introduce the parton construction of FQH wave functions, and then present our candidate wave function for the FQH state at $\nu = 5/2$. In the section that follows, we will analyze its properties.

\subsection{Background}
In the presence of a large magnetic field, interaction-induced mixing between Landau levels can be ignored. Relevant for this regime, the problem of interacting electrons restricted to any given Landau level is mathematically equivalent to the problem of electrons confined to the LLL, interacting via an effective ``pseudopotential''~\cite{Haldane83}. Throughout this work we shall therefore write down wave functions of electrons confined to the LLL, and assume the filled Landau levels below to be inert. Unless otherwise stated we shall assume the electrons to be fully spin polarized, and will ignore the effects of disorder and finite quantum well width.

The parton construction~\cite{Jain89b} provides a recipe for generating wave functions describing strongly correlated electrons in the LLL. Consider a system of $N$ electrons, with two-dimensional coordinates described by the complex numbers $z_j = x_j - i\, y_j$, where $j = 1 \ldots N$. The correlated $N$-particle wave function $\Psi_{\nu}(\{z_i\})$ is built out of a {\it product} of $k$ (uncorrelated) $N$-particle Slater determinant wave functions, $\Phi_{n_\alpha}(\{z_i\})$, with $\alpha = 1\ldots k$:
\begin{equation}
\label{eq:GeneralParton}\Psi_{\nu} = \mathcal{P}_{\rm LLL} \prod_{\alpha=1}^{k} \Phi_{n_{\alpha}}(\{z_{i}\}).
\end{equation}
Here $\mathcal{P}_{\rm LLL}$ projects the state to the LLL, and $\Phi_{n}(\{z_i\})$ is the wave function of $N$ electrons exactly filling $n$ Landau levels. We allow $n$ to take both positive and negative integer values; for $n < 0$, we define $\Phi_{n < 0} = [\Phi_{|n|}]^{*}$. Note that $k$ must be odd in order for $\Psi_\nu(\{z_i\})$ to be a valid fermionic wave function, antisymmetric under exchange of any two coordinates. 

The filling factor of the $k$-parton state $\Psi_\nu$ in Eq.~(\ref{eq:GeneralParton}) depends on both $k$ and $\{n_{\alpha}\}$ and is given by~\cite{Jain07}:
\begin{equation}
\label{eq:parton_filling}
\nu^{-1}=\sum_{\alpha=1}^{k} n^{-1}_{\alpha}.
\end{equation}
To produce the correct filling, each part $\Phi_{n_{\alpha}}$ must be constructed at a magnetic field  $B_{n_{\alpha}}=B\nu n^{-1}_{\alpha}$, where $B$ is the external magnetic field corresponding to filling factor $\nu$. 

To connect with familiar examples, we first note that the Laughlin wave function at $\nu=1/m$~\cite{Laughlin83} is an $m$-parton state, involving $m$ copies of $\Phi_1(\{z_i\}) = \prod_{i<j} (z_{i}-z_{j})e^{-\sum_{k} \frac{|z_k|^2}{4m\ell^2}}$:
\begin{equation}
\label{eq:Laughlin}\Psi^{\rm L}_{1/m} = e^{-\sum_{k} \frac{|z_k|^2}{4\ell^2}}\prod_{i<j} (z_{i}-z_{j})^{m} = \prod_{\alpha = 1}^m \Phi_{1}(\{z_{i}\}),
\end{equation}
where $\ell=\sqrt{\hbar c/(e B)}$ is the magnetic length. Below, we will follow the standard convention of omitting the ubiquitous Gaussian factors $e^{-\frac{1}{4\ell^2}\sum_{k}|z_{k}|^{2}}$. Composite fermion states~\cite{Jain89,Jain07} are also parton states: the composite fermion wave function at $\nu=n/(2pn\pm 1)$ (where $n$ and $p$ are positive integers), $\Psi^{\rm CF}_{n/(2pn\pm1)}$, is given by: 
\begin{equation}
\label{eq:CF} \Psi^{\rm CF}_{n/(2pn\pm1)} = \mathcal{P}_{\rm LLL} \Phi^{2p}_{1}\Phi_{\pm n}.
\end{equation}
This is a $(2p+1)$-parton state in which $2p$ partons are placed in $\nu=1$ IQH states, and a single parton forms an IQH state at $\nu=n$~\cite{footnote:reverse_flux}. 

The Laughlin and composite fermion states describe fractionalized phases in which the elementary excitations exhibit {\it Abelian} braiding statistics. Intriguingly, more general parton wave functions may describe {\it non-Abelian} phases. In particular, a parton wave function of the form:
\begin{equation}
\label{eq:non_Abelian_parton} 
\Psi_\nu = \Phi^{a}_{1}[\Phi_{\pm n}]^{b},
\end{equation}
with $n>1$, $a\geq 0$, and $b>1$, describes a non-Abelian state~\cite{Wen92b,Wen98,Wen99}. 
(Note that $k = a + b$ must be odd to ensure antisymmetry of the wave function, see above.)
An example of such a state is the Jain 221 state~\cite{Jain89b,Jain90,Wen91,Wu16}:
\begin{equation}
\label{eq:221}\Psi^{221}_{1/2}=\mathcal{P}_{\rm LLL}\Phi_{2}\Phi_{2}\Phi_{1}.
\end{equation} 
Recently, it has been proposed that this non-Abelian state may be realized in multi-layer graphene for a suitable set of parameters~\cite{Wu16}.

\subsection{The $\bar{2}\bar{2}111$ ansatz}
With the background in place, we now propose the following ``$\bar{2}\bar{2}111$'' parton state as a candidate ground state wave function at $\nu = 5/2$:
 \begin{equation}
  \Psi^{\bar{2}\bar{2}111}_{1/2}=\mathcal{P}_{\rm LLL}[\Phi_{2}]^{*}[\Phi_{2}]^{*}\Phi_{1}\Phi_{1}\Phi_{1}.
 \label{eq_parton_aPf}  
 \end{equation}
Comparing to the general form given in Eq.~(\ref{eq:non_Abelian_parton}), the $\bar{2}\bar{2}111$ state has $a = 3$, $n = 2$ (with the $-$ sign), and $b = 2$, and therefore characterizes a non-Abelian phase. Using Eq.~(\ref{eq:parton_filling}) it is easy to check that the state corresponds to a half-filled Landau level $\nu = 1/2$.

The $\bar{2}\bar{2}111$ wave function in Eq.~(\ref{eq_parton_aPf}) can be written in an alternative form, which is convenient for further manipulation. Returning to Eq.~(\ref{eq:CF}), note that the composite fermion wave function describing the state at $\nu = 2/3$ takes the form: $\Psi^{\rm CF}_{2/3} = \mathcal{P}_{\rm LLL} [\Phi_2]^{*}\Phi_1\Phi_1$~\cite{footnote:two_thirds_ph_Laughlin}.
Using this form, we write:  
\begin{equation}
\label{eq:22111_alternative} \Psi^{\bar{2}\bar{2}111}_{1/2} \sim \frac{[\Psi^{\rm CF}_{2/3}]^2}{\Phi_{1}}.
\end{equation}
In Eq.~(\ref{eq:22111_alternative}) we use $\sim$ because the right hand side of Eq.~(\ref{eq:22111_alternative}) differs slightly from the definition in Eq.~(\ref{eq_parton_aPf}) in the details of its projection to the LLL. We expect the details of projection to have only a minor effect on the properties of the state (see, e.g., Ref.~\onlinecite{Balram16b}). Note that it is possible to evaluate the wave function given in Eq.~(\ref{eq:22111_alternative}) for large systems using the Jain-Kamilla projection~\cite{Moller05,Davenport12,Balram15a} for the $2/3$ state. Therefore throughout this work we shall take $\Psi^{\bar{2}\bar{2}111}_{1/2} = [\Psi^{\rm CF}_{2/3}]^2/\Phi_{1}$. The state written in this form manifestly resides in the LLL and satisfies fermionic antisymmetry since $\Psi^{\rm CF}_{2/3}$ contains a factor of $\Phi_{1}$ in it. In the next paragraph we take a slight detour to discuss the spherical geometry, which will enable us to calculate many of the properties of our parton state. 

\subsection{Spherical geometry}
In the spherical geometry, $N$ electrons move on the surface of a sphere of radius $R=\sqrt{Q}\ell$, in the presence of a radial magnetic field generated by a monopole of strength $2Qhc/e$ located at the center of the sphere~\cite{Haldane83}. Due to the rotational symmetry of the system, the total orbital angular momentum $L$ and its $z$-component $L_{z}$ are good quantum numbers in this geometry. Uniform incompressible FQH states on the sphere have $L=0$ and occur at $2Q=\nu^{-1}N-\mathcal{S}$, where $\nu$ is the filling factor and $\mathcal{S}$ is a topological number called the shift~\cite{Wen92}. The state with $n$ filled Landau levels occurs at $2Q=(N-n^2)/n$ and thus has a shift of $\mathcal{S}=n$. Throughout this work we consider only uniform states, i.e., states with $L = L_{z} = 0$. States with $L>0$ (with $L$ scaling as $N$) are gapless in the thermodynamic limit. Hence we use $L = 0$ as a diagnostic to test if a state is compatible with remaining gapped in the thermodynamic limit (see Appendix~\ref{app:Ph-Pf}).

The wave functions defined on the two-dimensional plane can be translated into the spherical geometry by stereographic projection. For all the wave functions considered in this work this can be accomplished via the transformation $z_{i}-z_{j}\rightarrow u_{i}v_{j}-u_{j}v_{i}$, where $u_{i}=\cos(\theta_{i}/2)e^{i\phi_{i}/2}$ and $v_{i}=\sin(\theta_{i}/2)e^{-i\phi_{i}/2}$ are the spinor coordinates on the sphere. Here $\theta_i$ and $\phi_i$ are the polar and azimuthal angles locating particle $i$, respectively. 

In order to calculate overlaps of the parton state $\Psi^{\bar{2}\bar{2}111}_{1/2}$ with other candidate states, as well as its entanglement spectrum, it is useful to find its decomposition in the full Hilbert space of LLL Slater determinant states with $L_{z}=0$. To do so, we express the wave function as a linear combination of $L=0$ eigenstates as: $|\Psi_{\bar{2}\bar{2}111}\rangle=\sum_{i}c_{i}|\psi^{L=0}_{i}\rangle$, where each $|\psi^{L=0}_{i}\rangle$ has a known expansion in the full Hilbert space of Slater determinant states, $\{|\psi^{L_{z}=0}_{i}\rangle\}$. (Such a basis can be obtained by diagonalizing the $L^{2}$ operator or any spherically symmetric interaction in the full Hilbert space.) To obtain the coefficients $c_{i}$ we evaluate the wave function for many $N$-particle configurations on the sphere, and obtain a set of linear equations which we then invert to obtain $c_{i}$. This method works when the number of $L=0$ states is not very large, since for very large systems it is hard to find a set of well-conditioned linearly independent equations. Alternately one can use the method of Ref.~\onlinecite{Balram16b} to obtain $c_{i}$ using the Monte Carlo method by calculating overlaps with all LLL states. In this work we shall use the former method to evaluate the decomposition of $\Psi_{\bar{2}\bar{2}111}$ in the full Hilbert space for up to $N=10$ electrons, which suffices for our purposes.

\begin{table}
\centering
\begin{tabular}{|c|c|c|c|c|}
\hline
$N$ & $2Q$ & $|\langle \Psi^{1{\rm LL}}_{1/2}|\Psi^{\rm aPf}_{1/2} \rangle|$ &  $|\langle \Psi^{\bar{2}\bar{2}111}_{1/2}|\Psi^{\rm aPf}_{1/2} \rangle|$ & $|\langle \Psi^{1{\rm LL}}_{1/2}|\Psi^{\bar{2}\bar{2}111}_{1/2} \rangle|$  \\ \hline
4  & 9   		&0.8162    & 0.9639 &	0.9406	\\ \hline
6  & 13  		&0.8674    & 0.9686 &	0.9385	\\ \hline
8  & 17 	 	&0.8376    & 0.9523 &	0.9327	\\ \hline
10 & 21 	 	&0.8194    & 0.9397 & 	0.8975	\\ \hline
\end{tabular}
\caption{\label{overlaps_n_1_LL_2_3_square_over_1_parton} Overlaps of the ground state at the anti-Pfaffian flux in the $n=1$ Landau level (obtained by exact diagonalization), $\Psi^{1{\rm LL}}_{1/2}$, with the anti-Pfaffian, $\Psi^{\rm aPf}_{1/2}$, and $\Psi^{\bar{2}\bar{2}111}_{1/2}$ parton states. The numbers for $|\langle \Psi^{1{\rm LL}}_{1/2}|\Psi^{\rm aPf}_{1/2} \rangle|$ were previously given in Refs.~\onlinecite{Morf98,Scarola02,Pakrouski15}.}
\end{table}
\section{Properties of $\Psi^{\bar{2}\bar{2}111}_{1/2}$}\label{sec:candidate_parton}

\begin{figure*}[ht]
\begin{center}
\includegraphics[width=1.0\textwidth]{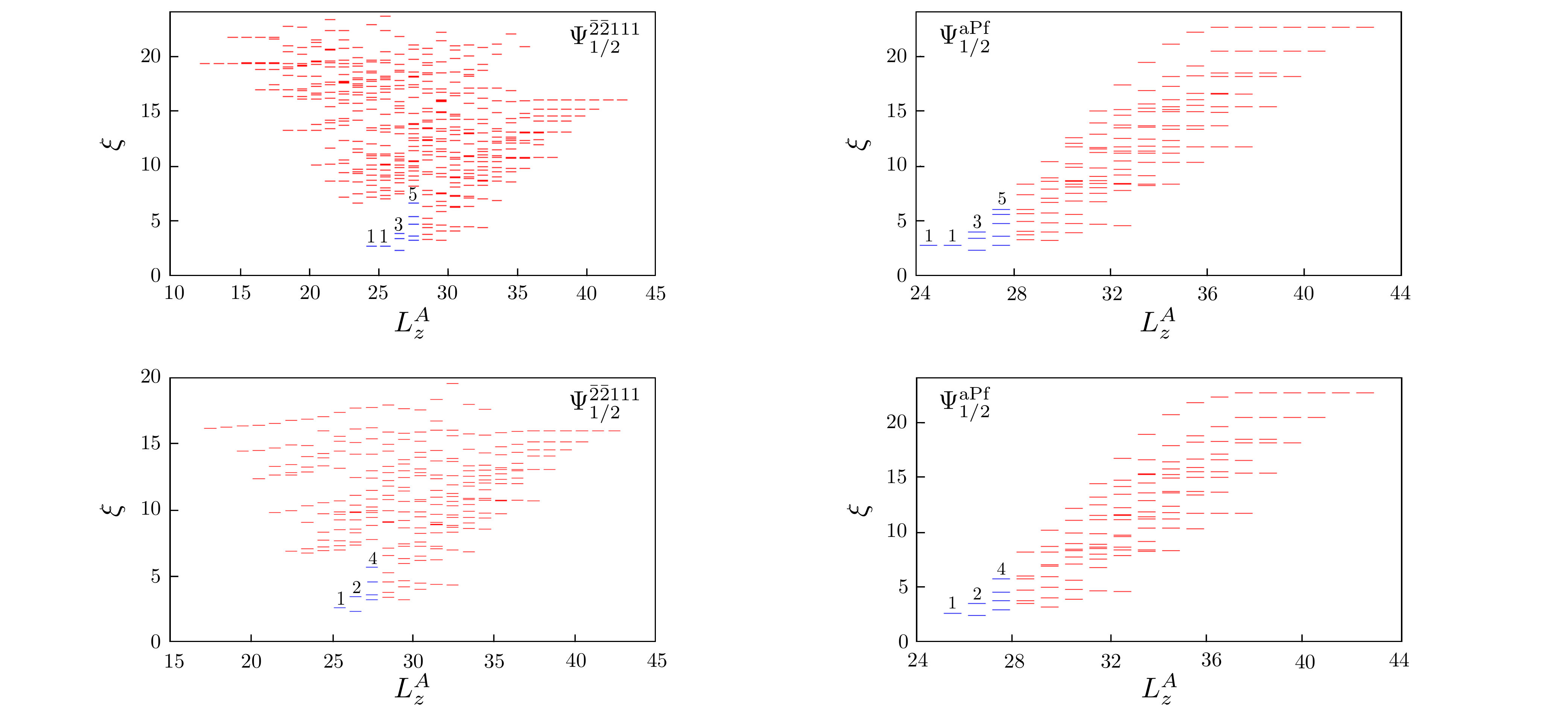}
\caption{(color online) Orbital entanglement spectrum of the $\Psi^{\bar{2}\bar{2}111}_{1/2}$ state for $N=10$ electrons at a flux $2Q=21$ on the sphere. The two subsystems $A$ and $B$ with respect to which the entanglement spectrum is calculated each have $N_{A}=N_{B}=5$ electrons and $l_{A}=l_{B}=11$ orbitals (top panels) and $l_{A}=12$ and $l_{B}=10$ orbitals (bottom panels). The entanglement levels are labeled by the $z$-component of the total orbital angular momentum of the $A$ subsystem, $L_{z}^{A}$. For comparison we also show the corresponding entanglement spectra for the anti-Pfaffian state, $\Psi^{\rm aPf}_{1/2}$ (right panels). The counting of low-lying levels (blue dashes) in the top panels, from $L_{z}^{A}=24$ (going from left to right) goes as $1,1,3,5,\cdots$, while in the bottom panels, from $L_{z}^{A}=25.5$ (going from left to right) goes as $1,2,4,\cdots$, and is identical for the states within each row.}
\label{fig:entanglement_spectra_parton}
\end{center}
\end{figure*}

We now analyze the properties of the $\bar{2}\bar{2}111$ state, and compare with other known candidates for the ground state at $\nu = 5/2$. We begin by briefly summarizing the properties of the Pfaffian and anti-Pfaffian reference states, and then show that the $\bar{2}\bar{2}111$ state is in the same phase as the anti-Pfaffian. 

The Moore-Read Pfaffian (Pf) state is described by the wave function~\cite{Moore91}:
\begin{equation}
\Psi^{\rm Pf}_{1/2}={\rm Pf}\left(\frac{1}{z_{i}-z_{j}}\right)\prod_{i<j}(z_{i}-z_{j})^{2},
\end{equation}
where ${\rm Pf}$ is the Pfaffian of a skew-symmetric matrix~\cite{footnote:Pfaffian}. The state $\Psi^{\rm Pf}_{1/2}$ represents a topologically non-trivial $p$-wave paired condensate of fully spin polarized composite fermions~\cite{Read00}. Important characteristics of the Pfaffian state are that in the spherical geometry it occurs at a shift of $\mathcal{S}=3$, it carries a thermal Hall conductance of $\kappa_{xy}=3/2~(\pi^2 k_{\rm B}^{2}T/(3h))$, and has a six-fold topological degeneracy when realized on a torus~\cite{Read00,Peterson08,Pakrouski15}. 

Because $\Psi^{\rm Pf}_{1/2}$ describes a half-filled Landau level, its particle-hole conjugate $\Psi^{\rm aPf}_{1/2}$, dubbed the anti-Pfaffian (aPf), also occurs at $\nu = 1/2$~\cite{Levin07,Lee07}. The anti-Pfaffian is topologically distinct from the Pfaffian: it has a shift of $\mathcal{S}=-1$ on the sphere, and carries a thermal Hall conductance of $\kappa_{xy}=-1/2~(\pi^2 k_{\rm B}^{2}T/(3h))$. Following recent measurements of the thermal Hall conductance on the $\nu = 5/2$ FQH plateau~\cite{Banerjee17b}, there has been considerable interest in a particle-hole symmetric variant of the Pfaffian state. However, the proposed candidate wave function for this state~\cite{Zucker16} does not appear to have good overlap with the numerically-exact ground state for any reasonable two-body interaction (see Appendix~\ref{app:Ph-Pf}).

To compare, the $\bar{2}\bar{2}111$ parton state occurs at a shift of $\mathcal{S}=-1$, which is the same as the anti-Pfaffian shift [see Eq.~(\ref{eq:parton_extended_flux_particle_relation}) below]. Furthermore, for small system sizes where it can be checked, $\Psi^{\bar{2}\bar{2}111}_{1/2}$ has a good overlap with the anti-Pfaffian state as well as with the numerically-exact ground state obtained using the second Landau level Coulomb pseudopotentials, $\Psi^{1{\rm LL}}_{1/2}$, see Table \ref{overlaps_n_1_LL_2_3_square_over_1_parton}. These features indicate that in the thermodynamic limit $\Psi^{\bar{2}\bar{2}11}_{1/2}$ will be in the same phase as the anti-Pfaffian.

We obtain further strong evidence linking $\Psi^{\bar{2}\bar{2}111}_{1/2}$ to the anti-Pfaffian phase by comparing the entanglement spectra of the two states. The entanglement spectrum has been useful in characterizing many FQH states, as the counting of low-lying entanglement levels provides a fingerprint of the topological order of the state~\cite{Li08}. In Fig.~\ref{fig:entanglement_spectra_parton} we show the orbital entanglement spectrum~\cite{Haque07} of the $\Psi^{\bar{2}\bar{2}111}_{1/2}$ state for a system of $N=10$ electrons on a sphere with flux $2Q=21$. The multiplicities of the low-lying entanglement levels of $\Psi^{\bar{2}\bar{2}111}_{1/2}$ are identical to those of the anti-Pfaffian. Thus we conclude that $\Psi^{\bar{2}\bar{2}111}_{1/2}$ lies in the anti-Pfaffian phase.

We now give an analytical argument to explicitly show that $\Psi^{\bar{2}\bar{2}111}_{1/2}$ lies in the anti-Pfaffian phase. 
The wave function in Eq.~(\ref{eq_parton_aPf}) (ignoring the projection to the LLL) can be re-written as:
\begin{equation}
\label{eq:Parton2}\Psi^{\bar{2}\bar{2}111}_{1/2} = \Psi_{1/3}[\Phi^{2}_{2}]^{*}.
\end{equation}
Viewed in this way, we may express the state in Eq.~(\ref{eq:Parton2}) in terms of partons $\wp = f b$, comprised of a fermion $f=f_{1}f_{2}f_{3}$ that forms a $\nu=1/3$ Laughlin state (with each $f_{i}$ forming a $\nu=1$ state), and a boson $b$ that forms the state described by the symmetric wave function $[\Phi^{2}_{2}]^{*}$. 

Let us then first understand the properties of the symmetric wave function $\Phi^{2}_{2}$. This state itself can be understood
in terms of a parton construction with $b = f_4 f_5$, where $f_4$ and $f_5$ each form a $\nu = 2$ IQH state. A lattice version of this state was studied explicitly in ~\cite{Zhang13a}. The state's topological order can be read off using techniques discussed in Ref. ~\cite{Wen99,Barkeshli10} (see Appendix~\ref{app:Phi2square}); $[\Phi^{2}_{2}]^*$ describes a non-Abelian FQH state with central charge $c = -5/2$, whose fusion rules coincide with the fusion rules of quasiparticles in the Ising topological quantum field theory. There is precisely one such topological quantum field theory~\cite{Kitaev06,Rowell09}. 

An alternative way to realize the unique topological order with $c = -5/2$ and Ising fusion rules is in terms of the composite fermion construction. Consider the state where we attach one unit of flux to $b$, and the resulting composite fermion forms a paired state with odd angular momentum $l = -3$. In other words, we consider a parton construction where $b = \psi_{-1} \psi_{l=-3}$, where $\psi_{-1}$ is a fermion in a $\nu = -1$ state and $\psi_{l=-3}$ is a fermion in a $l = -3$ paired state. This theory has a central charge $c = -1 - 3/2 = -5/2$. It is straightforward to verify that all such composite fermion states with odd angular momentum pairing have the same Ising fusion rules, due to the Majorana zero mode localized at vortex cores of odd angular momentum paired states. 

Comparing the forms above, the topological order of the original state $\wp = f b$ can be understood by rewriting the parton construction as:
\begin{equation}
\wp = f b = f_1 f_2 f_3 \psi_{-1} \psi_{l=-3} = a f_1 f_2 \psi_{l=-3}  .
\end{equation}
Here $a = f_3 \psi_{-1}$ is a boson, and it forms a superfluid state (the product of two fermionic states with $\nu=1$ and $\nu=-1$ is a bosonic superfluid 
state ~\cite{Barkeshli14,Shapourian16}). Since the boson $a$ is trivially condensed, it does not contribute to the topological order of $\wp$. This allows us to ignore $a$, simplifying our parton construction to:
\begin{equation}
\wp = f_1 f_2 \psi_{l=-3}, 
\end{equation}
where $f_1$ and $f_2$ are in $\nu=1$ state, and $\psi_{l=-3}$ describes an angular momentum $l = -3$ paired state~\cite{footnote:Pfaffian_as_p_wave_paired_state}. Such a state is known to describe the anti-Pfaffian state~\cite{Barkeshli15}. One can easily check that the central charge of this state is $c = 1 - 3/2 = -1/2$, which matches the anti-Pfaffian value~\cite{Levin07,Lee07}. We remark here that a similar construction shows that the Jain 221 state~\cite{Jain89b,Jain90,Wu16} can be thought of as an $l = +3$ paired state of composite fermions.

\section{Applications and extensions}\label{sec:applicatons_extensions}
\begin{figure}
  \begin{center}
    \includegraphics[width=\columnwidth]{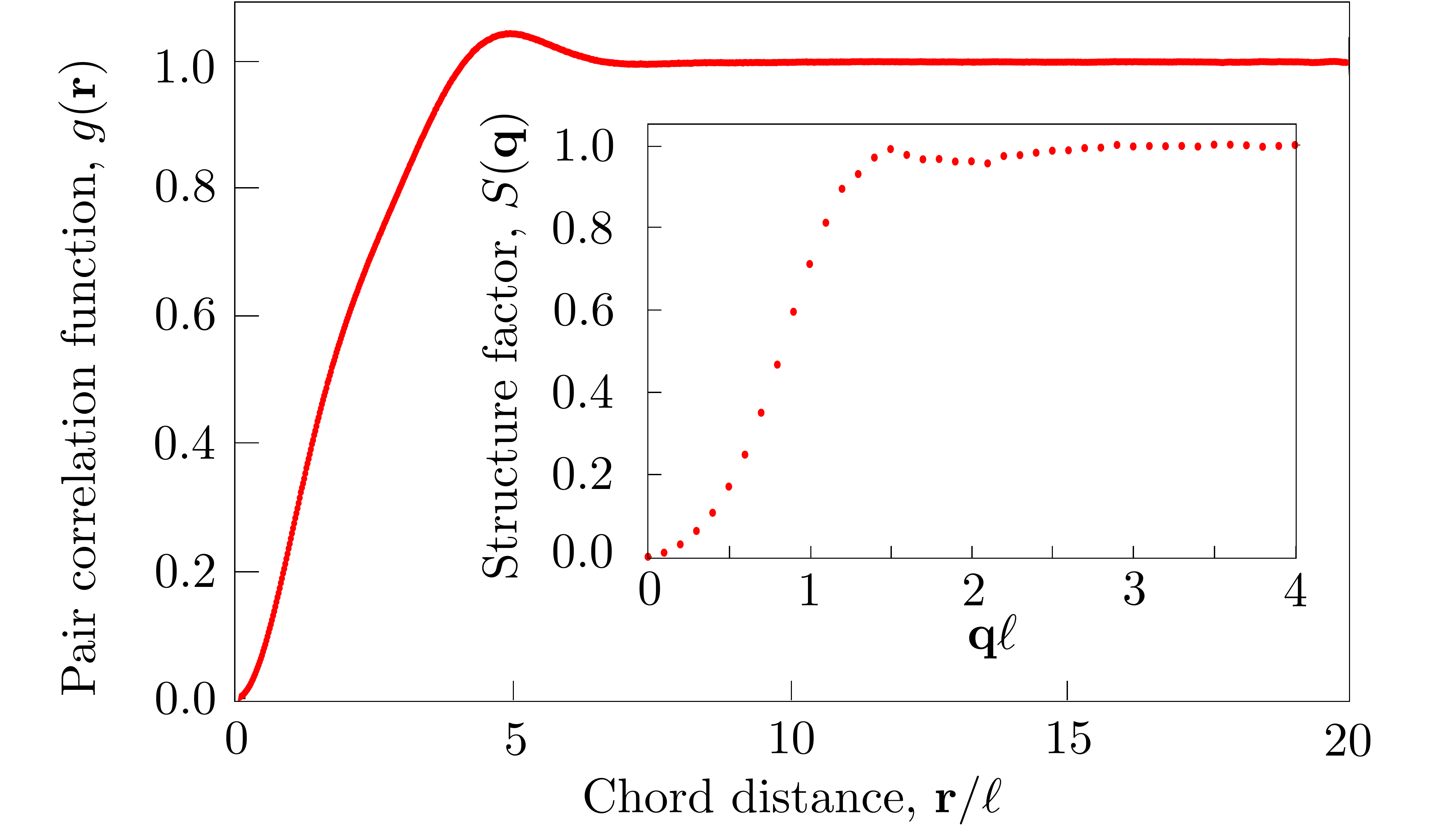}
    \caption{(color online) Pair-correlation function $g({\bf r})$, where ${\bf r}$ is the chord distance between electrons on the sphere, and its Fourier transform, the structure factor $S({\bf q})$ (inset) in the state $\Psi^{\bar{2}\bar{2}111}_{1/2}$. The calculation is performed for $N=100$ electrons on the sphere, with flux $2Q=201$. The pair-correlation function shows a ``shoulder''-like feature at intermediate distances which is typical of non-Abelian states.}
    \label{fig:pair_correlation_structure_factor}
  \end{center}
\end{figure}

The utility of the parton wave function that we have presented in Eq.~(\ref{eq:22111_alternative}) is that it can be constructed for systems much larger than those accessible via brute force particle-hole conjugation of the Pfaffian or by exact diagonalization. As a proof of principle, in Fig.~\ref{fig:pair_correlation_structure_factor} we show the pair-correlation function and static structure factor for a system of $N=100$ electrons on the sphere obtained using the Monte Carlo method described in detail in Ref.~\onlinecite{Balram17}. We see a ``shoulder''-like bump at short distances in the pair-correlation function which is considered to be a characteristic of non-Abelian states that involve clustering~\cite{Moore91,Read99,Rezayi09,Bonderson12,Hutasoit16}. 

Generalizing the parton ansatz in Eq.~(\ref{eq_parton_aPf}), we may construct a family of wave functions of the form:
\begin{equation}
\label{eq:parton_extended} \Psi_{\frac{n}{pn\pm k}}=\mathcal{P}_{\rm LLL}\Phi_{1}^{p}[\Phi_{\pm n}]^{k}.
\end{equation} 
In the spherical geometry the wave function in Eq.~(\ref{eq:parton_extended}) occurs at flux~\cite{Jain07}: 
\begin{equation}
\label{eq:parton_extended_flux_particle_relation}
2Q=p(N-1)\pm k\frac{(N-n^{2})}{n}=\frac{pn\pm k}{n}N \mp kn-p,
\end{equation}
corresponding to the filling factor $\nu=n/(pn\pm k)$ and a shift $\mathcal{S}=p\pm kn$. 
We have the following four cases to consider for the symmetry of the wave function:
\begin{itemize}
 \item \emph{$p$ odd and $k$ odd}: wave function is symmetric and represents a state of bosons
 \item \emph{$p$ odd and $k$ even}: wave function is anti-symmetric and represents a state of fermions
 \item \emph{$p$ even and $k$ odd}: wave function is anti-symmetric and represents a state of fermions
 \item \emph{$p$ even and $k$ even}: wave function is symmetric and represents a state of bosons
\end{itemize}
For even $p$ and $k=1$ these wave functions form the series of composite fermion states~\cite{Jain89,Jain07}, at $\nu=n/(pn\pm 1)$. 
The symmetry (bosonic or fermionic) of any of these wave functions can be changed by adding a Pfaffian or an anti-Pfaffian factor, and the anti-Pfaffian in turn could appropriately be represented by a parton wave function using our construction. With the Pfaffian factor, the $k=1$ and $p=3$ series gives Bonderson-Slingerland states~\cite{Bonderson08}. The properties of other members of this family, including extensions to multicomponent systems (involving spin, valley, or orbital indices) remain to be explored.

\section{Discussion}\label{sec:discussion}

In this work we have introduced a parton wave function, denoted by $\Psi^{\bar{2}\bar{2}111}_{1/2}$, describing a FQH state in a half-filled Landau level that lies in the same phase as the anti-Pfaffian state. For the system sizes we could study, the overlap of $\Psi^{\bar{2}\bar{2}111}_{1/2}$ with the numerically-exact ground state is slightly better than the corresponding overlap of $\Psi^{\rm aPf}_{1/2}$ with the exact ground state. In contrast to the anti-Pfaffian wave function, for which no simple expression is known in first quantized form, the $\Psi^{\bar{2}\bar{2}111}_{1/2}$ wave function is straightforward to evaluate. Thus the parton ansatz enables numerical studies of the anti-Pfaffian phase on much bigger systems than were possible previously. Furthermore, the large system sizes accessible may make it possible to do a Berry phase calculation to demonstrate non-Abelian braiding statistics of the exotic quasiparticles hosted by this state.

As a proof of concept, we demonstrated the utility of the parton ansatz by computing the pair correlation function and the static structure factor for a relatively large system, containing 100 electrons on the sphere. Looking ahead, the $\Psi^{\bar{2}\bar{2}111}_{1/2}$ ansatz could be used, for example, to obtain the real-space entanglement spectrum~\cite{Sterdyniak12,Dubail12} of the anti-Pfaffian phase for a large system. Another application is to use the parton state to fix the phase of the anti-Pfaffian state in a fixed-phase diffusion Monte Carlo (DMC) calculation~\cite{Ortiz93,Zhang16,footnote:DMC_LL_mixing}. Such a calculation would allow one to determine whether the Pfaffian or anti-Pfaffian state is preferred in the presence of LL mixing~\cite{Bishara09,Wang09,Wojs10,Storni10,Rezayi11,Peterson14,Zaletel15,Pakrouski15,Tyler15,Rezayi17}, in the thermodynamic limit. It will be interesting to explore whether other unexplained, experimentally-observed FQH states can be described by the general family of parton states.

\begin{acknowledgments}
We gratefully acknowledge useful discussions with Michael Levin, Zlatko Papi\'c, Steve Simon, Ady Stern, Nicolas Regnault and Jainendra Jain. The Center for Quantum Devices is funded by the Danish National Research Foundation. This work was supported by the European Research Council (ERC) under the European Union Horizon 2020 Research and Innovation Programme, Grant Agreement No. 678862 and the Villum Foundation. MB is supported by NSF CAREER (DMR-1753240) and JQI-PFC-UMD. Some of the numerical calculations were performed using the DiagHam package, for which we are grateful to its authors. 
\end{acknowledgments}

\appendix
\section{Particle-hole symmetric Pfaffian}
\label{app:Ph-Pf}
A half-filled state which is particle-hole symmetric must occur at flux $2Q=2N-1$ (on the sphere), and carry a thermal Hall conductance of $\kappa_{xy}=1/2~(\pi^2 k_{\rm B}^{2}T/(3h))$. Recent experiments measuring the thermal Hall conductance at $\nu=5/2$ are consistent with this value~\cite{Banerjee17b}. The following candidate wave function for such a state, dubbed the particle-hole symmetric Pfaffian (PH-Pf), has been proposed~\cite{Zucker16, footnote:PHPf_no_P_LLL}:
\begin{equation}
\Psi^{\rm PH\text{-}Pf}_{1/2}=\mathcal{P}_{\rm LLL}~\left[{\rm Pf}\left(\frac{1}{z_{i}-z_{j}}\right)\right]^{*}\prod_{i<j}(z_{i}-z_{j})^{2}.
\label{eq_wf_PH_Pf}
\end{equation}
This state has a high overlap with its particle-hole conjugate, but a low overlap with the lowest energy $L = 0$ state obtained by exact diagonalization using the second LL Coulomb pseudopotential (see Table~\ref{tab:overlaps_Pf_bar}). 
For the ideal second LL Coulomb pseudopotentials the ground state at $2Q=2N-1$ does not always have $L=0$ for different numbers of particles, $N$ (see entries indicated by daggers in Table \ref{tab:overlaps_Pf_bar}). Hence it is unlikely that the state will remain gapped in the thermodynamic limit (see main text).
However, it is possible that effects such as Landau level mixing, disorder~\cite{Mross17,Wang17a,Lian18}, and finite width corrections may stabilize a gap at this flux value~\cite{Zucker16}. Recently, Simon~\cite{Simon18} has proposed that the thermal Hall measurements would be consistent with the anti-Pfaffian if its Majorana edge mode does not thermally equilibrate with the bosonic edge modes. 

We observe that the overlap of $\Psi^{\rm PH\text{-}Pf}_{1/2}$ with the lowest energy $L=0$ state of the \emph{lowest} LL Coulomb interaction is fairly good (see Table \ref{tab:overlaps_Pf_bar}); we speculate that this could possibly describe a particle-hole symmetric pairing instability of the composite fermion Fermi sea.

We mention here that Jolicoeur~\cite{Jolicoeur07} constructed a wave function which differs from the state shown in Eq.~(\ref{eq_wf_PH_Pf}) by a factor of $\prod_{i<j}|z_{i}-z_{j}|^{2}$ before projection. We have checked using the Monte Carlo method that after projection, for small values of $N\leq 10$, these two states have high overlap with each other indicating that they are similar in nature~\cite{Regnault18}. For $N=12$ electrons the overlap between these two states was found to be 0.993 by Mishmash \emph{et al.}~\cite{Mishmash18}. A state consisting of alternating stripes of Pfaffian and anti-Pfaffian has also been put forth as a viable candidate state for $\nu=5/2$~\cite{Wan16}.  
\begin{table}
\centering
\begin{tabular}{|c|c|c|c|c|c|}
\hline
$N$ & $2Q$ & $|\langle \mathcal{C}_{\rm ph}\Psi^{\rm PH\text{-}Pf}_{1/2}|\Psi^{\rm PH\text{-}Pf}_{1/2} \rangle|$ & $|\langle \Psi^{1{\rm LL}}_{1/2}|\Psi^{\rm PH\text{-}Pf}_{1/2} \rangle|$ & $|\langle \Psi^{0{\rm LL}}_{1/2}|\Psi^{\rm PH\text{-}Pf}_{1/2} \rangle|$  \\ \hline
6  & 11  		&0.999(1) &0.001(1) 		& 0.990(1)		\\ \hline
8  & 15 	 	&0.999(1) &0.002(1) 		& 0.975(1)$^{\dagger}$	\\ \hline
10 & 19 	 	&0.998(5) &0.002(1)$^{\dagger}$ & 0.938(6)$^{\dagger}$ 	\\ \hline
\end{tabular}
\caption{\label{tab:overlaps_Pf_bar} Overlaps of the PH-Pfaffian trial state $\Psi^{\rm PH\text{-}Pf}_{1/2}$, Eq.~(\ref{eq_wf_PH_Pf}), with its particle-hole conjugate ($\mathcal{C}_{\rm ph}\Psi^{\rm PH\text{-}Pf}_{1/2}$),  as well as the lowest energy $L=0$ states found by exact diagonalization using the Coulomb pseudopotentials for the second Landau level ($\Psi^{1{\rm LL}}_{1/2}$) and the lowest Landau level ($\Psi^{0{\rm LL}}_{1/2}$). Daggers indicate cases where the ground state does not have $L=0$. The projection to the lowest Landau level is carried out by calculating the overlap of the state $\left[{\rm Pf}\left( [z_{i}-z_{j}]^{-1}\right)\right]^{*}\prod_{i<j}(z_{i}-z_{j})^{2}$ with each basis function in the lowest Landau level using the Monte Carlo method of Ref.~\onlinecite{Balram16b}. 
Statistical errors are reported in parentheses. }
\end{table}

\subsection{Model interactions to search for the particle-hole symmetric Pfaffian state at $\nu=1/2$} 
As mentioned above, for the second Landau level Coulomb pseudopotentials, our results indicate that it is likely that the ground state
at the particle-hole symmetric flux, $2Q=2N-1$, will become gapless in the thermodynamic limit.
We investigate a wider set of two-body pseudopotentials, and ask if any of them give rise to a uniform ground state at this flux. Since the wave function of Eq.~(\ref{eq_wf_PH_Pf}) is particle-hole symmetric to a very high degree, it is at least feasible that it may be realized as the ground state of a simple two-body interaction potential. For this purpose we have considered two kinds of interactions which will be described below.

\subsubsection{Generic pairwise interactions with non-negative, nonvanishing values only for $V_{1}$, $V_{3}$ and $V_{5}$ } 
\label{subsubsec:pair_model}
We consider a set of model Hamiltonians with pairwise interactions of the form:
\begin{equation}
H=\sum_{i<j}\sum_{m}\mathcal{P}^{i,j}_{m}V_{m},  
\end{equation}
where $\mathcal{P}^{i,j}_{m}$ projects onto the subspace of two electrons, $i$ and $j$, with relative angular momentum $m$, and $V_{m}$ is the interaction energy in the relative angular momentum $m$ channel. We limit our search to Hamiltonians with $V_{m}=0$ for $m>5$. 

The overall scale of the pseudopotentials is irrelevant for determining the nature of the ground state wave function. (The energies and gaps do, of course, depend on the scale.) We therefore choose the normalization $V_1+V_3+V_5=1$. This choice then leaves only two independent parameters for the model interaction, and allows us to show results in a convenient form using a set of triangular maps. A given point on or inside the triangle corresponds to a particular ratio between $V_1$, $V_3$, and $V_5$. It is easiest to visualize the triangle as made up of the following three regions:
\begin{itemize}
 \item each vertex of the triangle corresponds to a single positive pseudopotential: $V_{1}=1$ (bottom left vertex), $V_{3}=1$ (bottom right vertex) and $V_{5}=1$ (top vertex).
 \item each edge of the triangle corresponds to two positive pseudopotentials $V_{m_{1}}=V$, $V_{m_{2}}=1-V$ with $0\leq V\leq 1$.
 \item each point inside the triangle corresponds to all three pseudopotentials being non-zero, with the size of a particular pseudopotential proportional to the distance from its respective vertex, in particular the centroid of the triangle has the values $(V_1,V_3,V_5)=(1/3,1/3,1/3)$.
\end{itemize}
The region outside of the triangle is not of our interest since it has at least one negative pseudopotential. 

\begin{figure*}[h]
\begin{center}
\fbox{\includegraphics[width=0.3\textwidth]{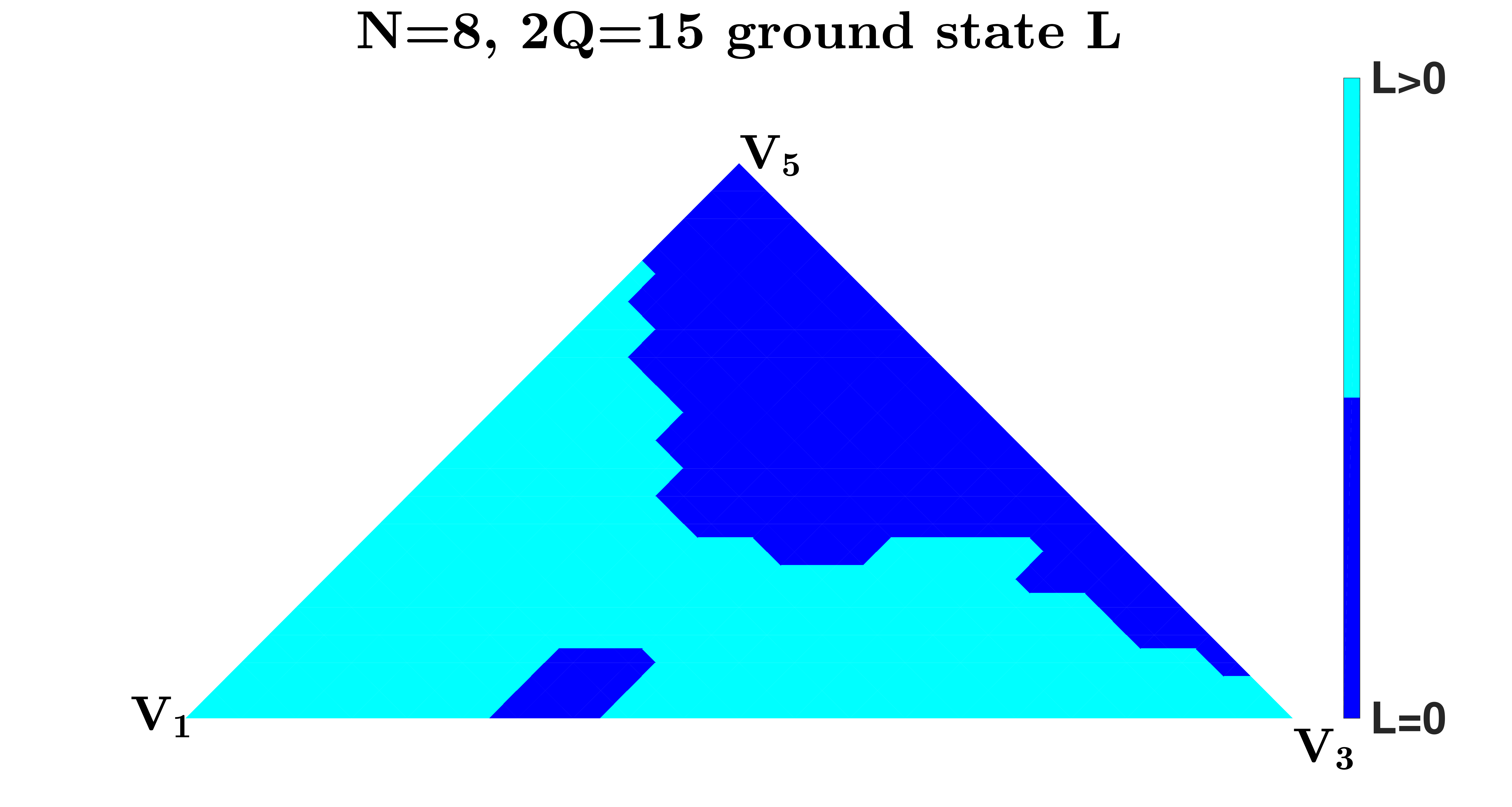}   }
\fbox{\includegraphics[width=0.3\textwidth]{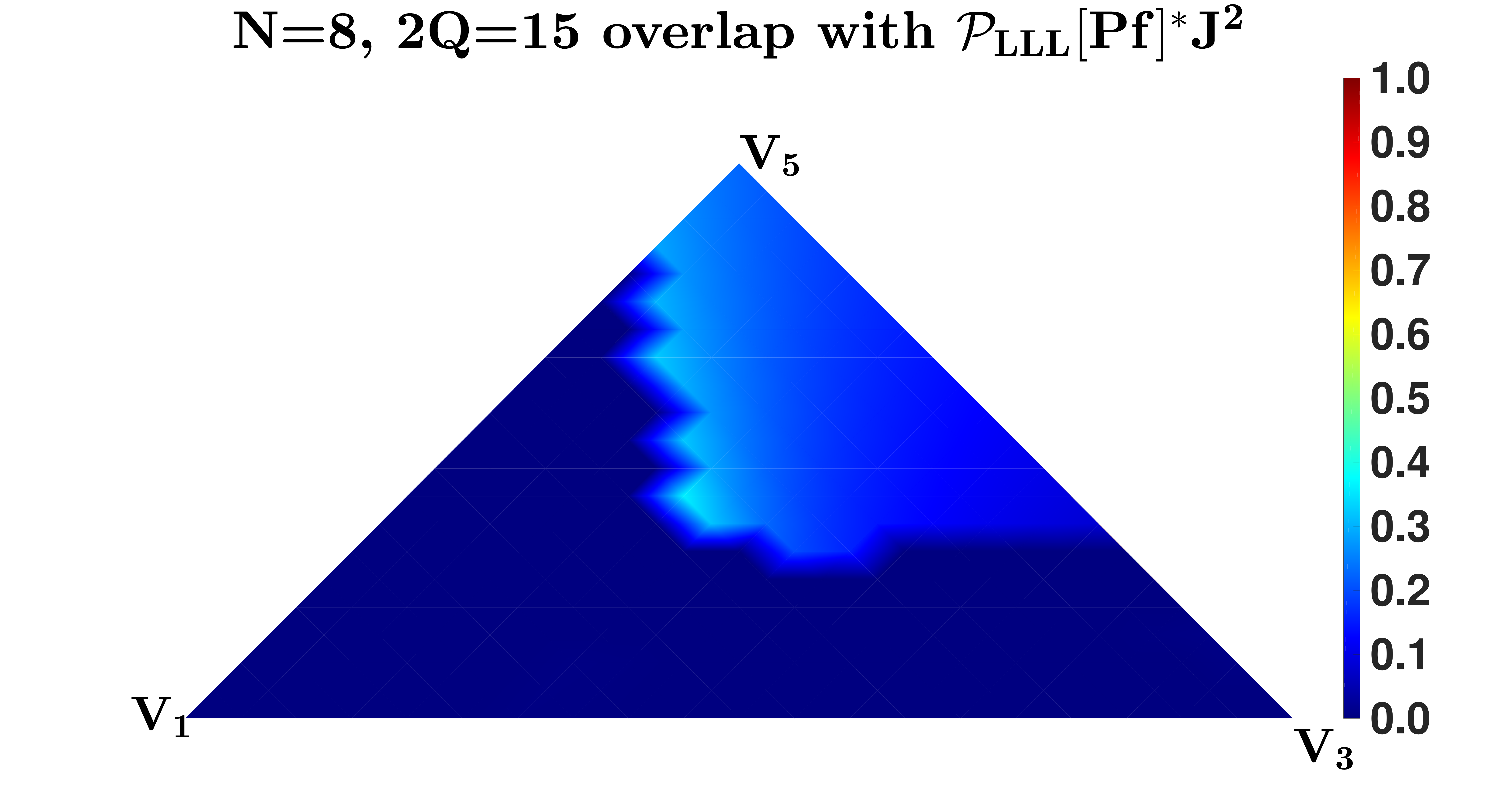} }  \\
\fbox{\includegraphics[width=0.3\textwidth]{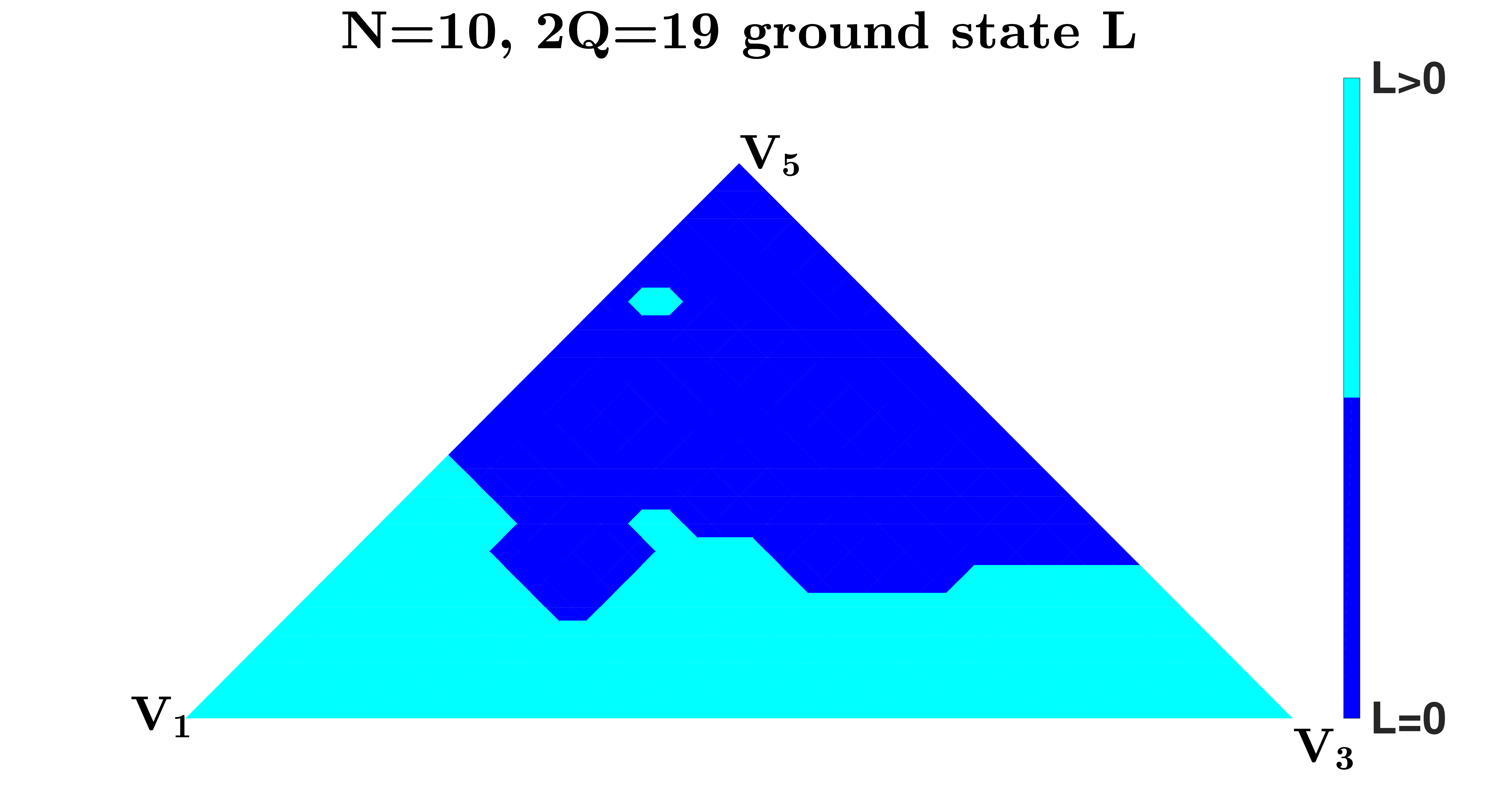}  }
\fbox{\includegraphics[width=0.3\textwidth]{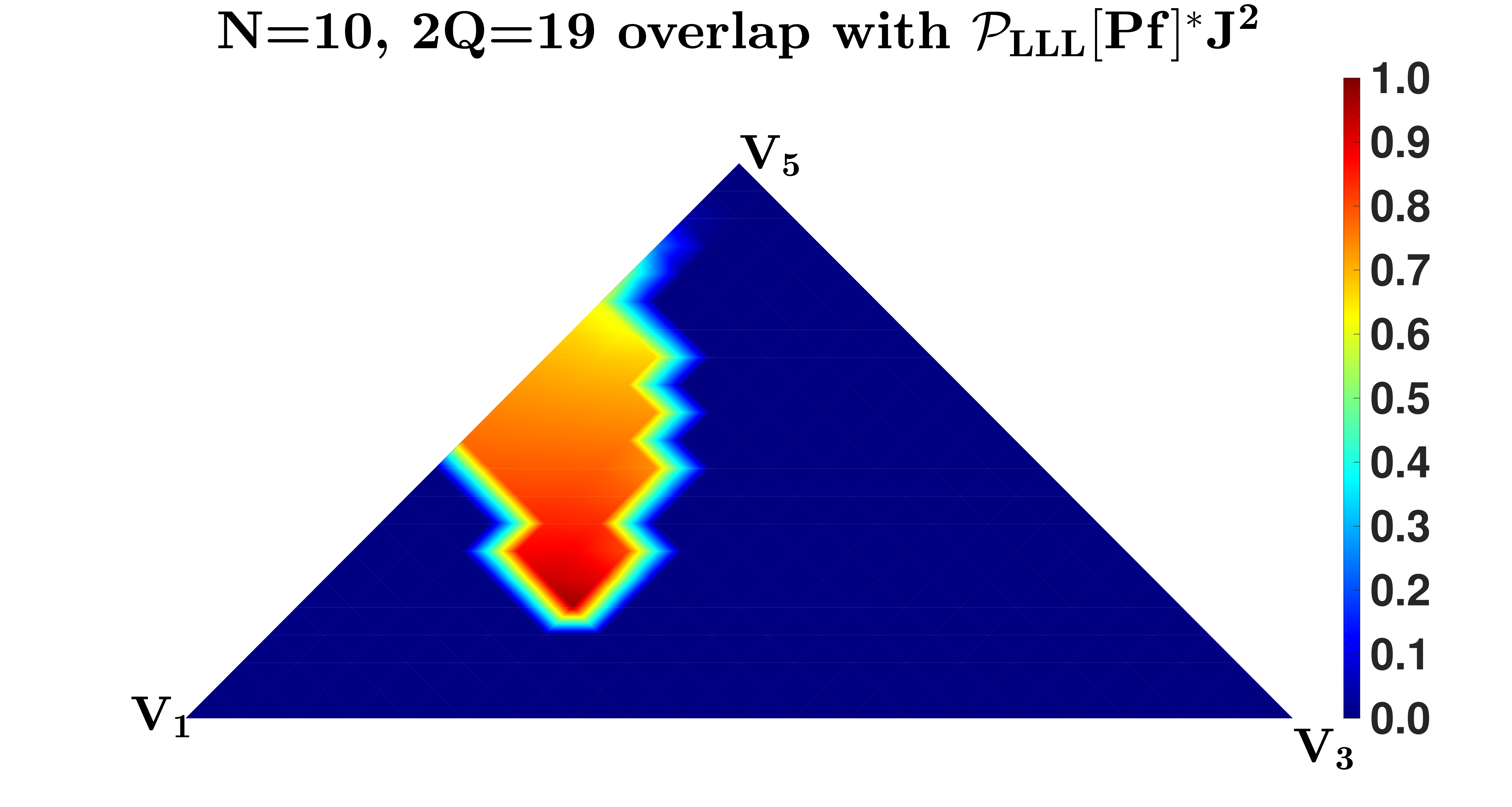} } \\
\fbox{\includegraphics[width=0.3\textwidth]{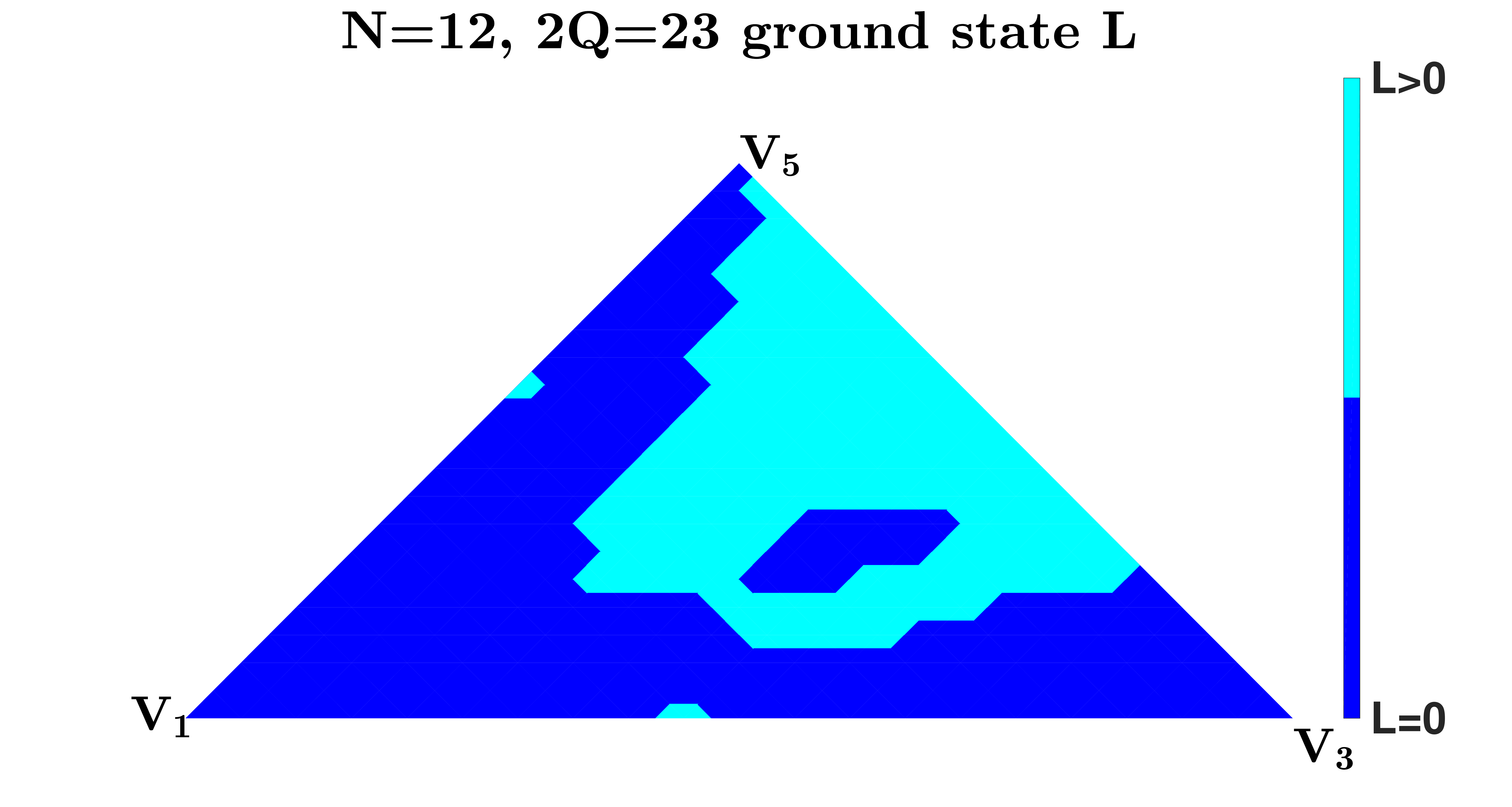}  }
\fbox{\includegraphics[width=0.3\textwidth]{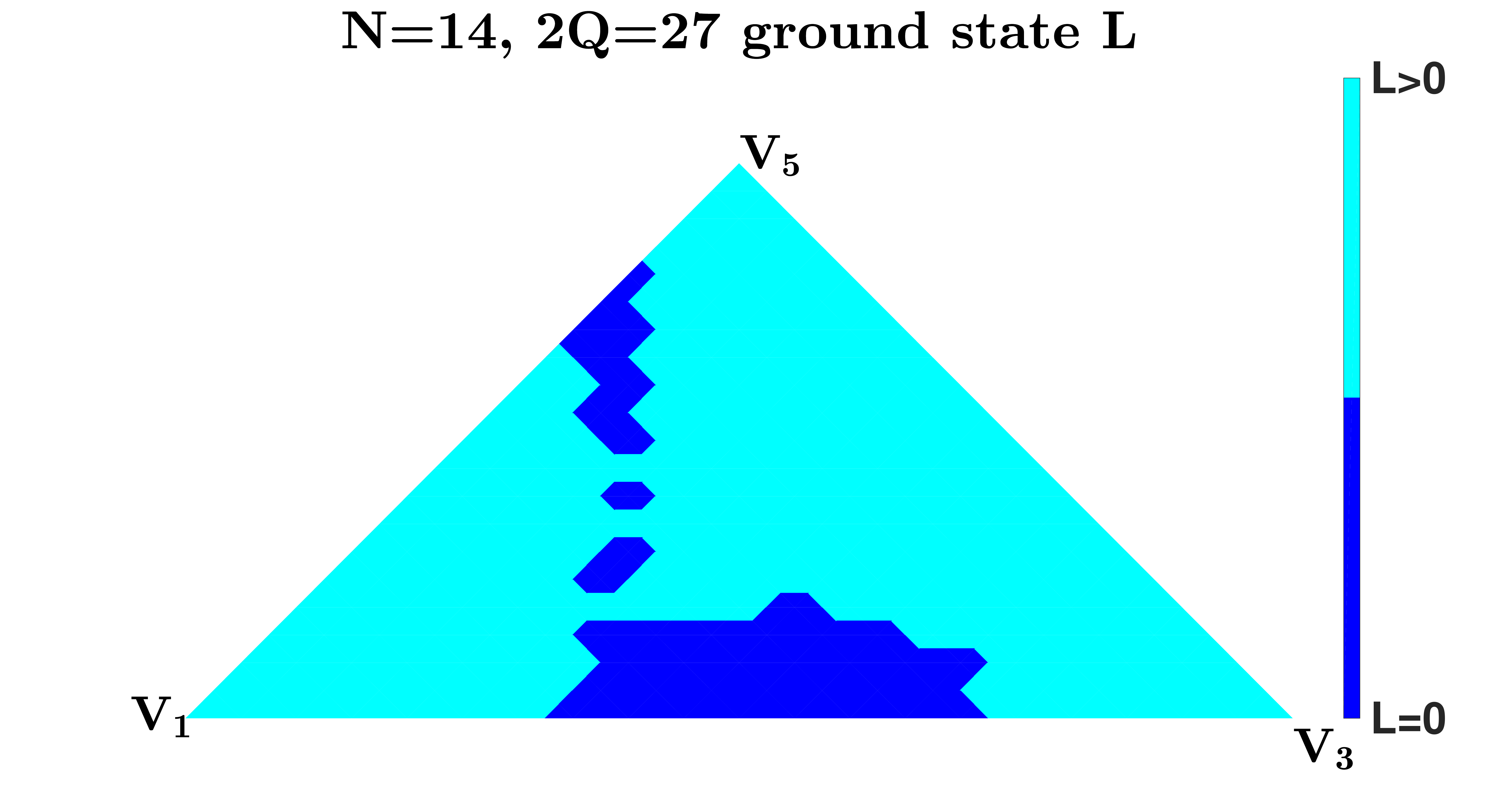}  }
\fbox{\includegraphics[width=0.3\textwidth]{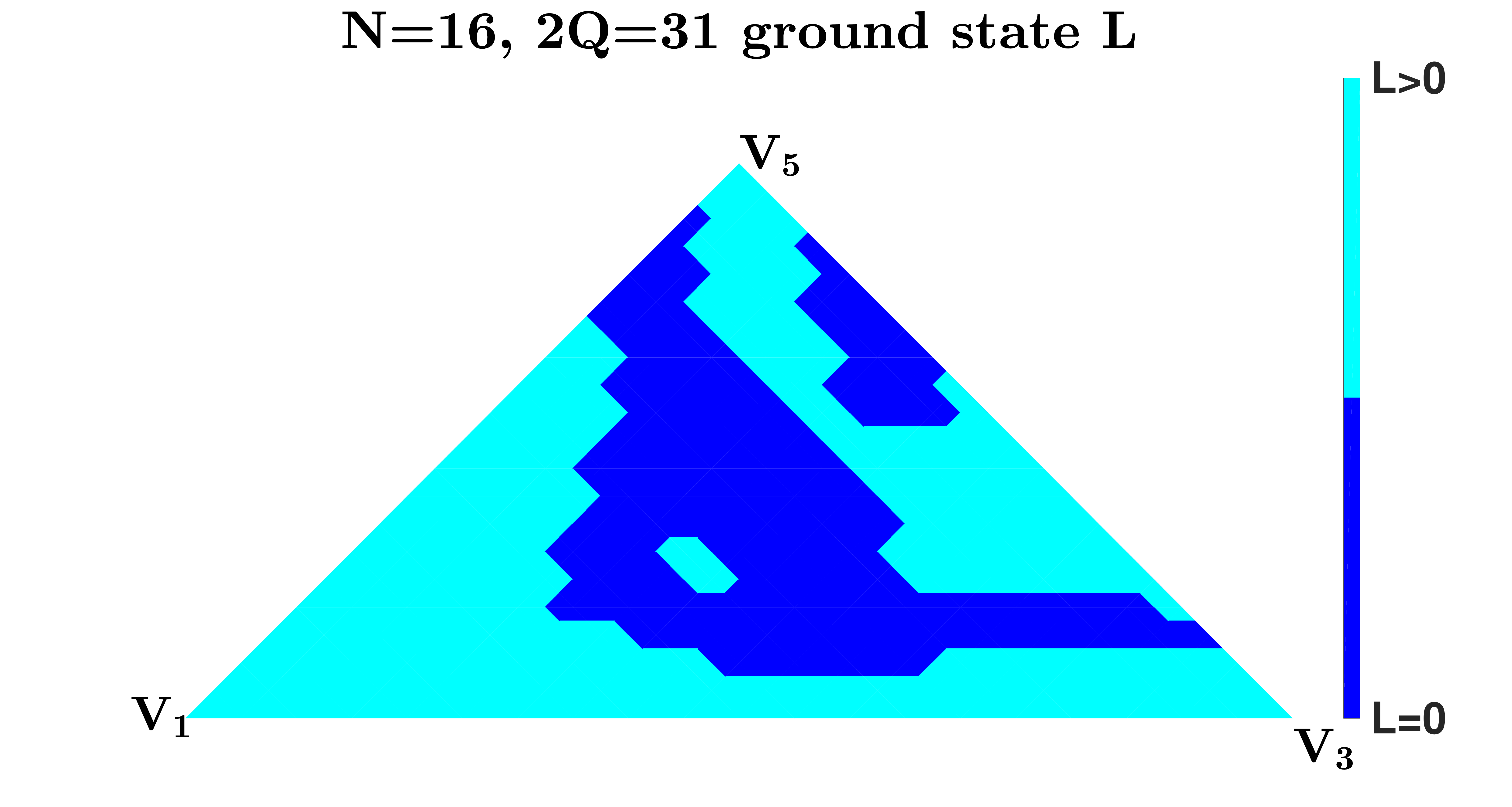}  }
\caption{(color online) The ground state orbital angular momentum, $L$, obtained in the spherical geometry using exact diagonalization at the particle-hole symmetric Pfaffian flux for a model interaction (see Sec.~\ref{subsubsec:pair_model} for a description of the interaction). Regions of parameter space where the ground state is uniform (i.e., has $L = 0$), and therefore compatible with maintaining a finite energy gap in the thermodynamic limit, are indicated in dark blue shade. In the right panels of the top two rows we show the overlaps of the particle-hole symmetric Pfaffian trial state $\mathcal{P}_{\rm LLL}[{\rm Pf}]^{*}J^{2}=\mathcal{P}_{\rm LLL}~\left[{\rm Pf}\left(\{(z_{i}-z_{j})^{-1}\}\right)\right]^{*}\prod_{i<j}(z_{i}-z_{j})^{2} $ with the ground state of the model interaction, for $N=8$ and $N=10$ particles.}
\label{fig:spectra_model_pps_phPf}
\end{center}
\end{figure*}

In Fig.~\ref{fig:spectra_model_pps_phPf} we show plots of the ground state total orbital angular momentum $L$ (we only distinguish if the state has $L=0$ or $L > 0$, where the latter is a heuristic indicator of gaplessness in the thermodynamic limit), and the overlap with the candidate PH-Pf state at the particle-hole symmetric flux for different numbers of electrons. We do not find any region of parameter space where the ground state consistently occurs at $L = 0$ for all system sizes. Furthermore, in the region where the ground state is uniform for a particular $N$ (and hence could be compatible with maintaining a gap in the thermodynamic limit), we do not find a high overlap between the PH-Pf state and the ground state of the model interaction, for all system sizes.

\subsubsection{Interactions perturbed around the lowest and second Landau level Coulomb points}
\label{subsubsec:Coulomb_model}
We now consider interactions that are perturbed around the lowest and second Landau level Coulomb points. These interactions are more realistic than the ones considered in the previous section. We perturb the short-range part of the Coulomb interaction by changing the values of $V_{1}$ and $V_{3}$ to $V^{\rm Coulomb}_{1}+\delta V_{1}$ and $V^{\rm Coulomb}_{3}+\delta V_{3}$ respectively. 

\begin{figure*}[ht]
\begin{center}
\includegraphics[width=0.3\textwidth]{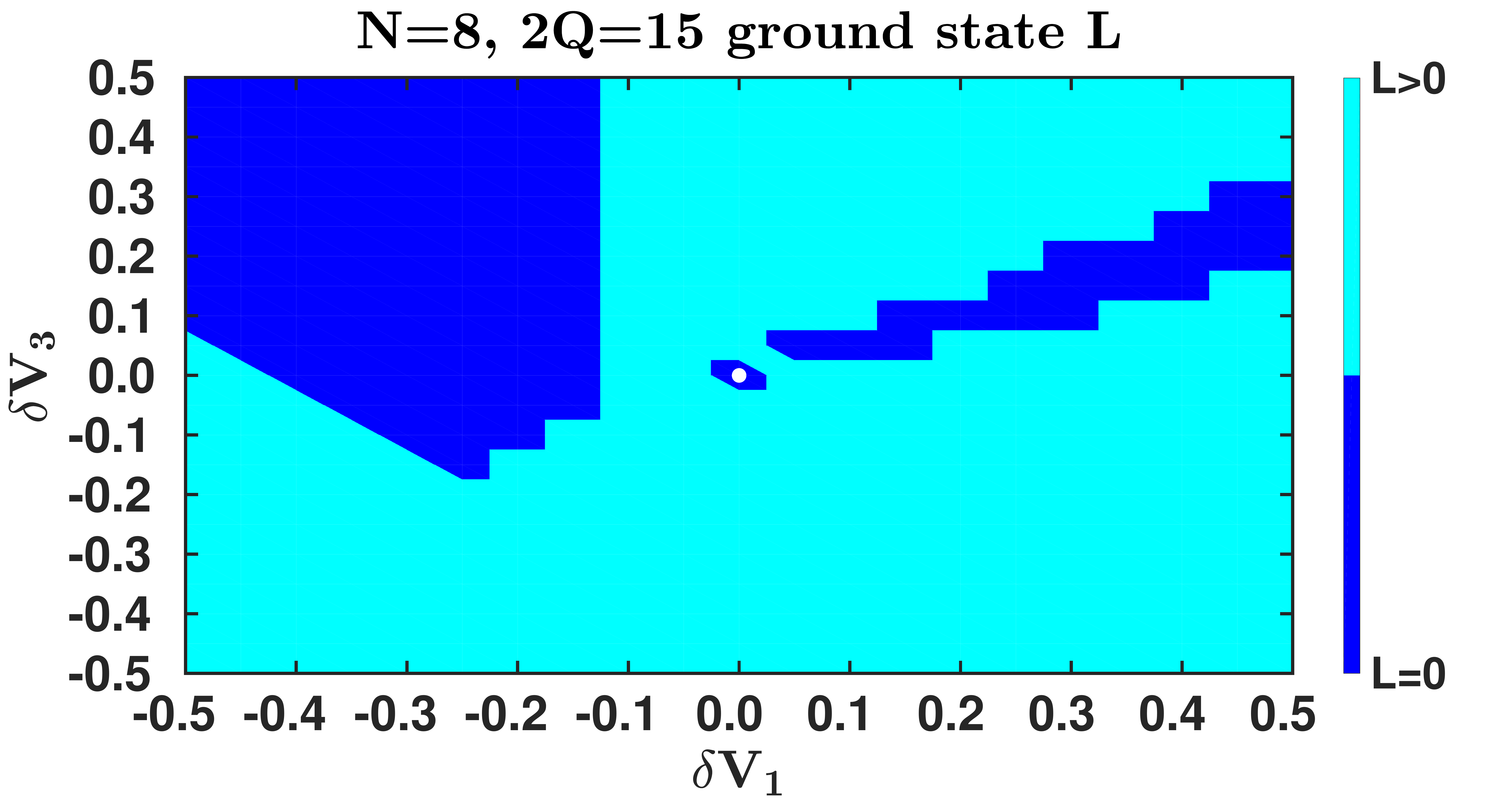} 
\includegraphics[width=0.3\textwidth]{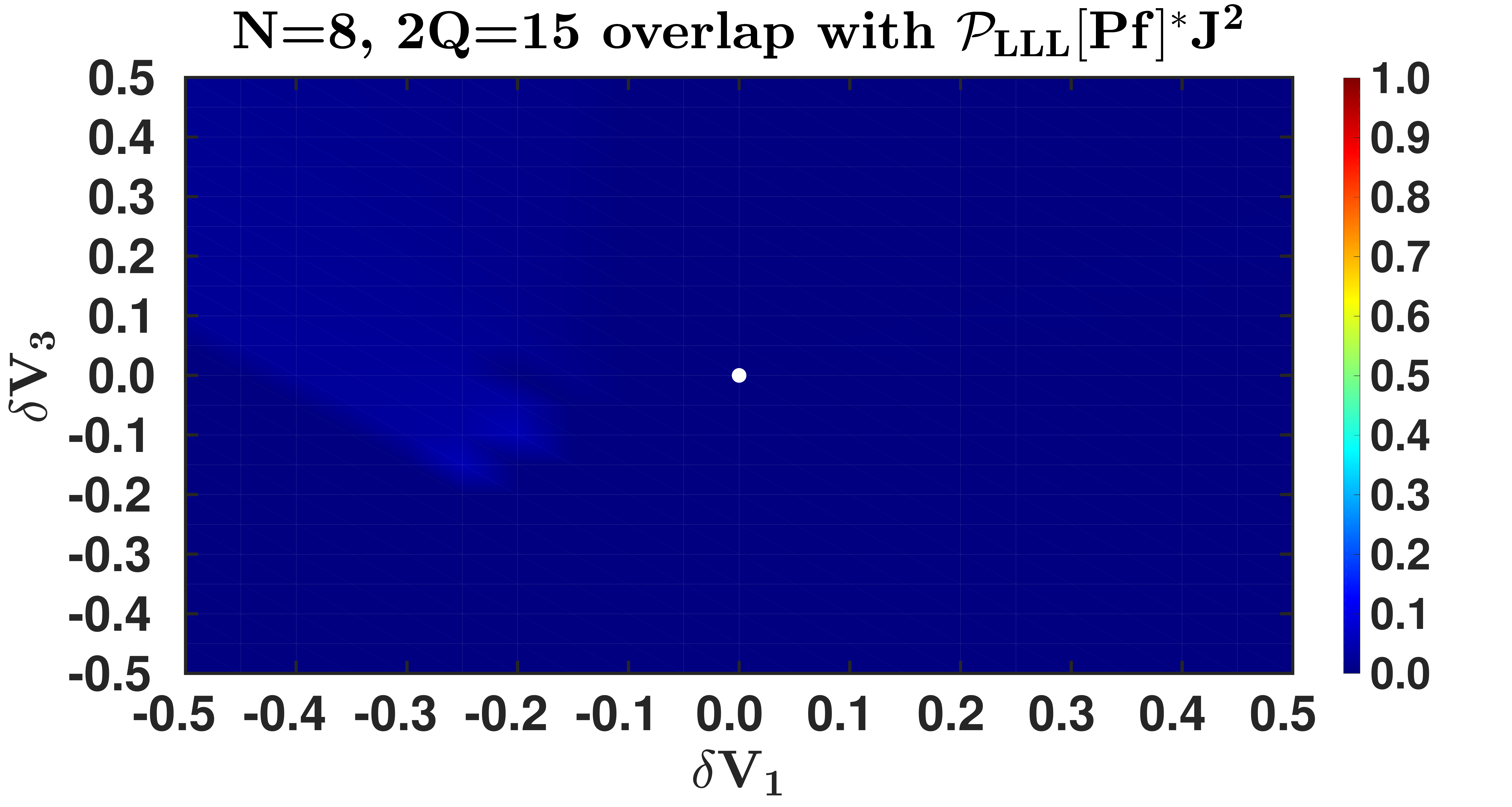}  \\
\includegraphics[width=0.3\textwidth]{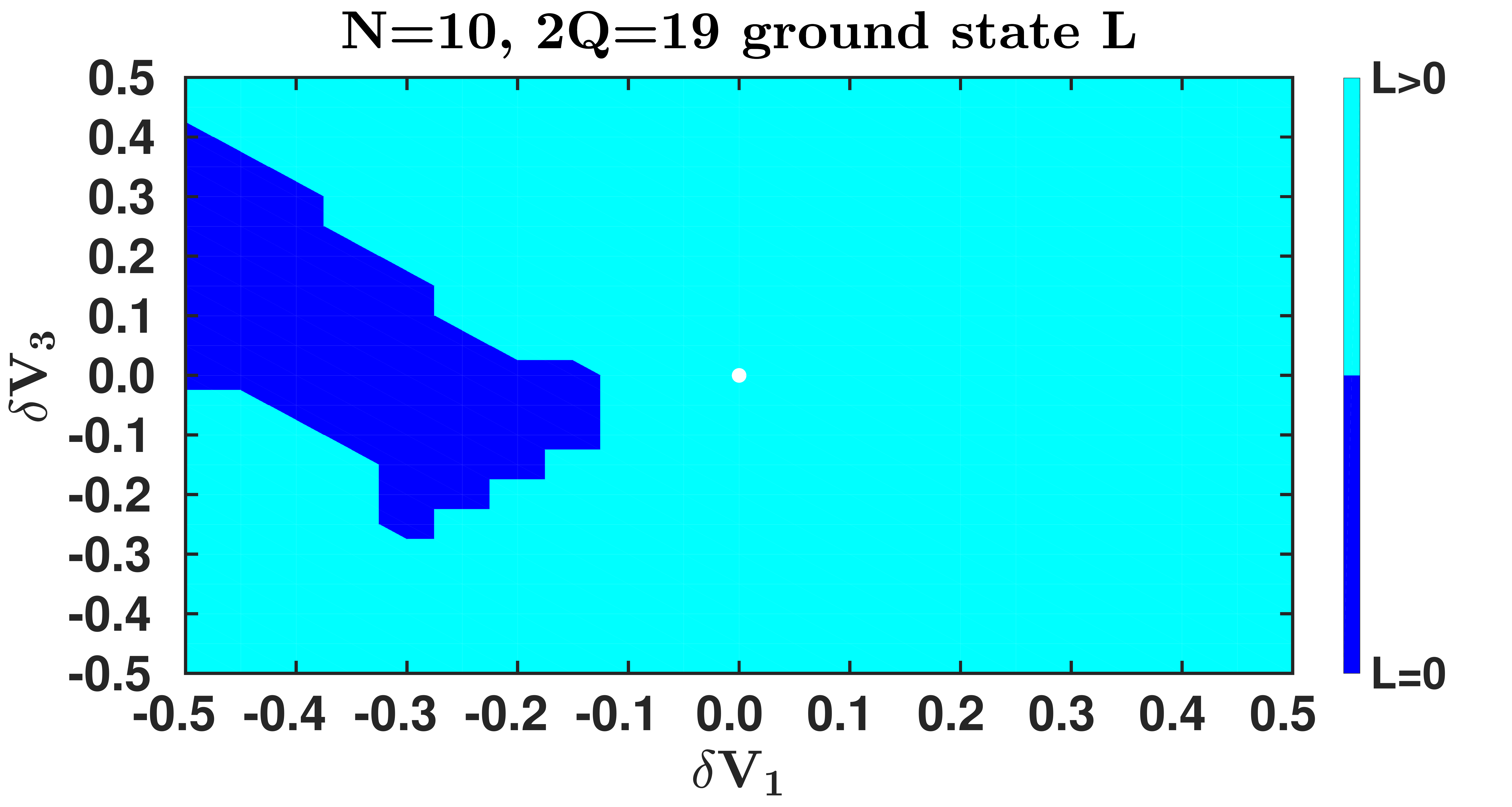} 
\includegraphics[width=0.3\textwidth]{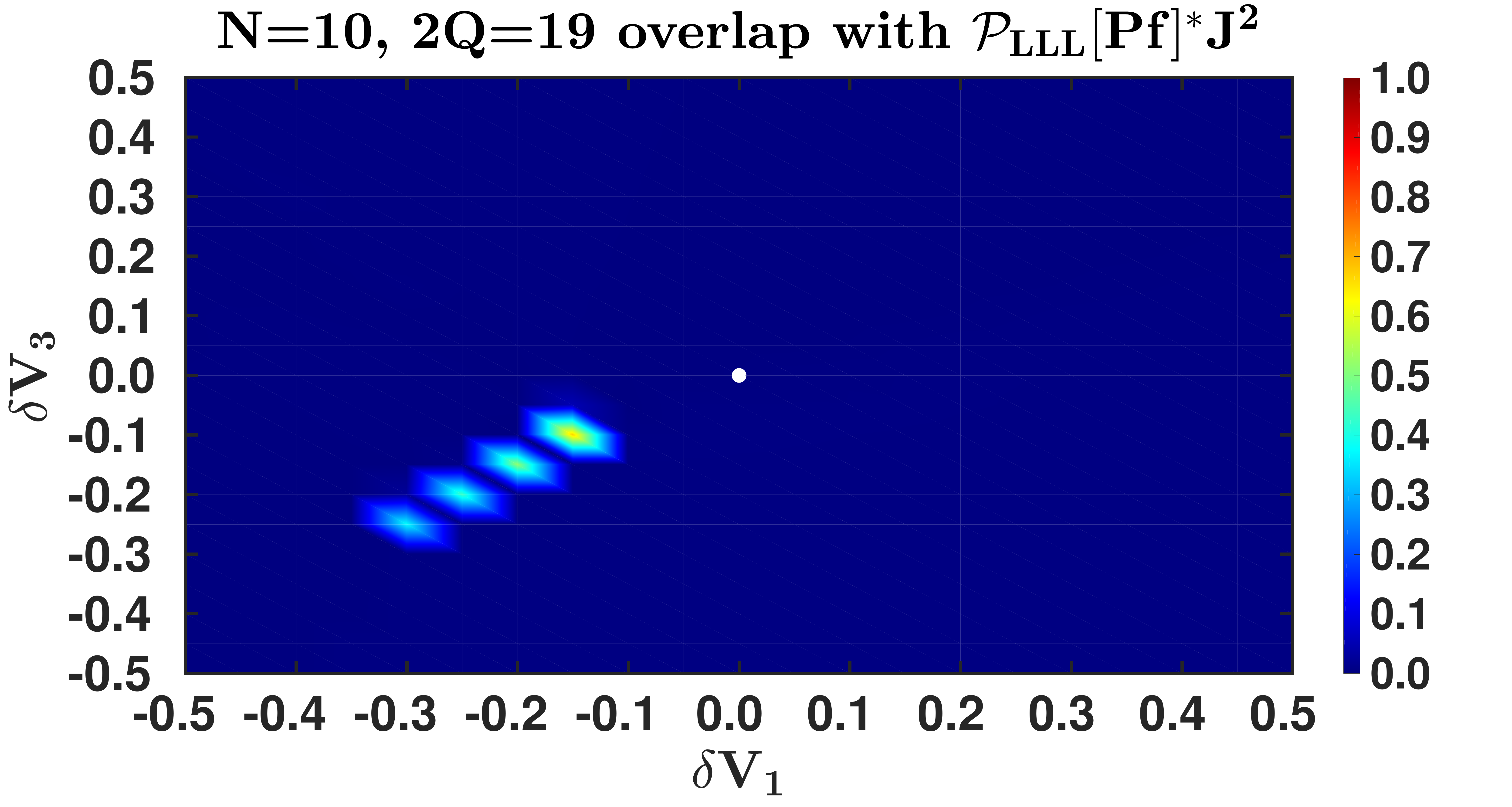} \\ 
\includegraphics[width=0.3\textwidth]{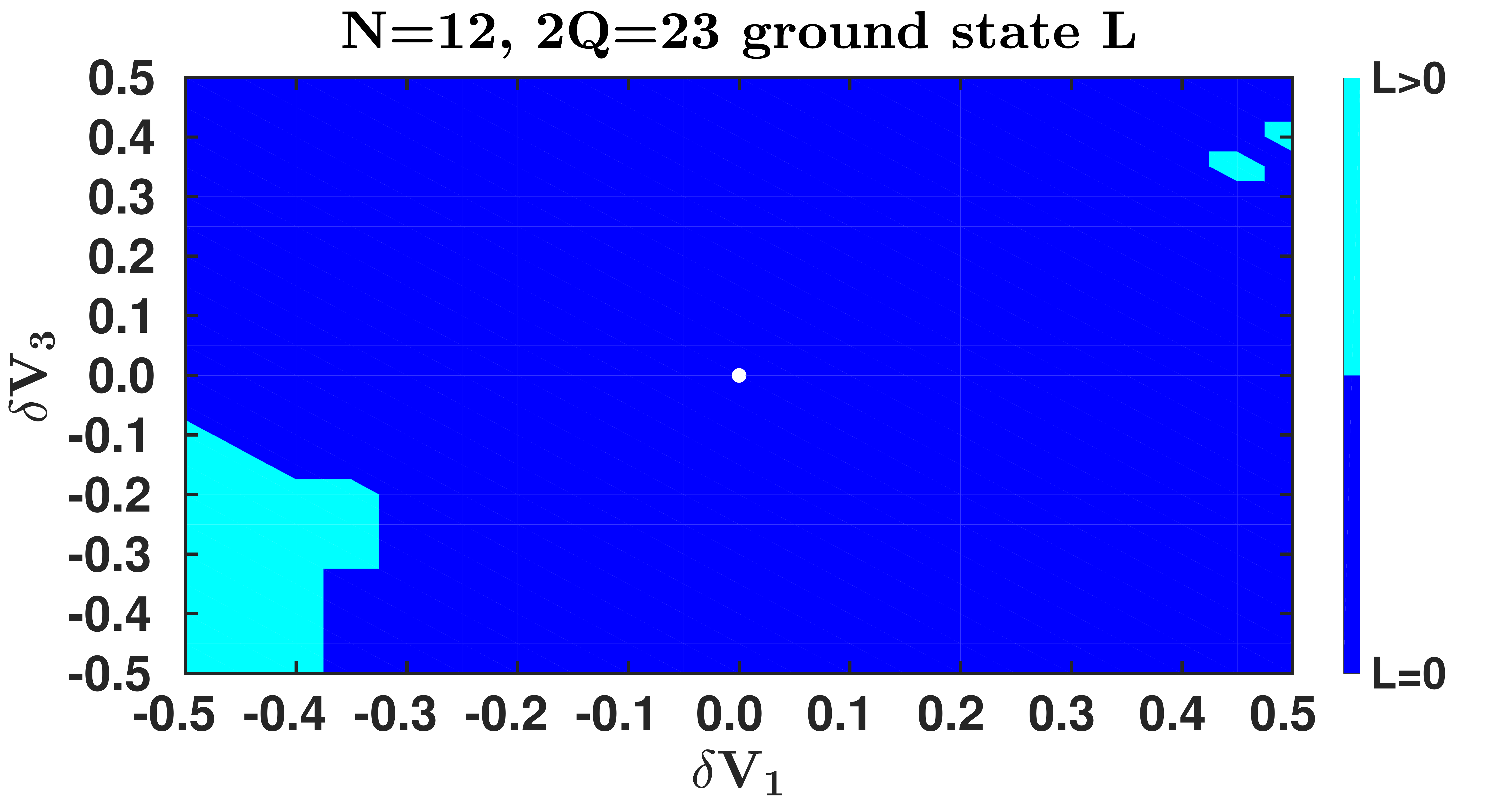}   
\includegraphics[width=0.3\textwidth]{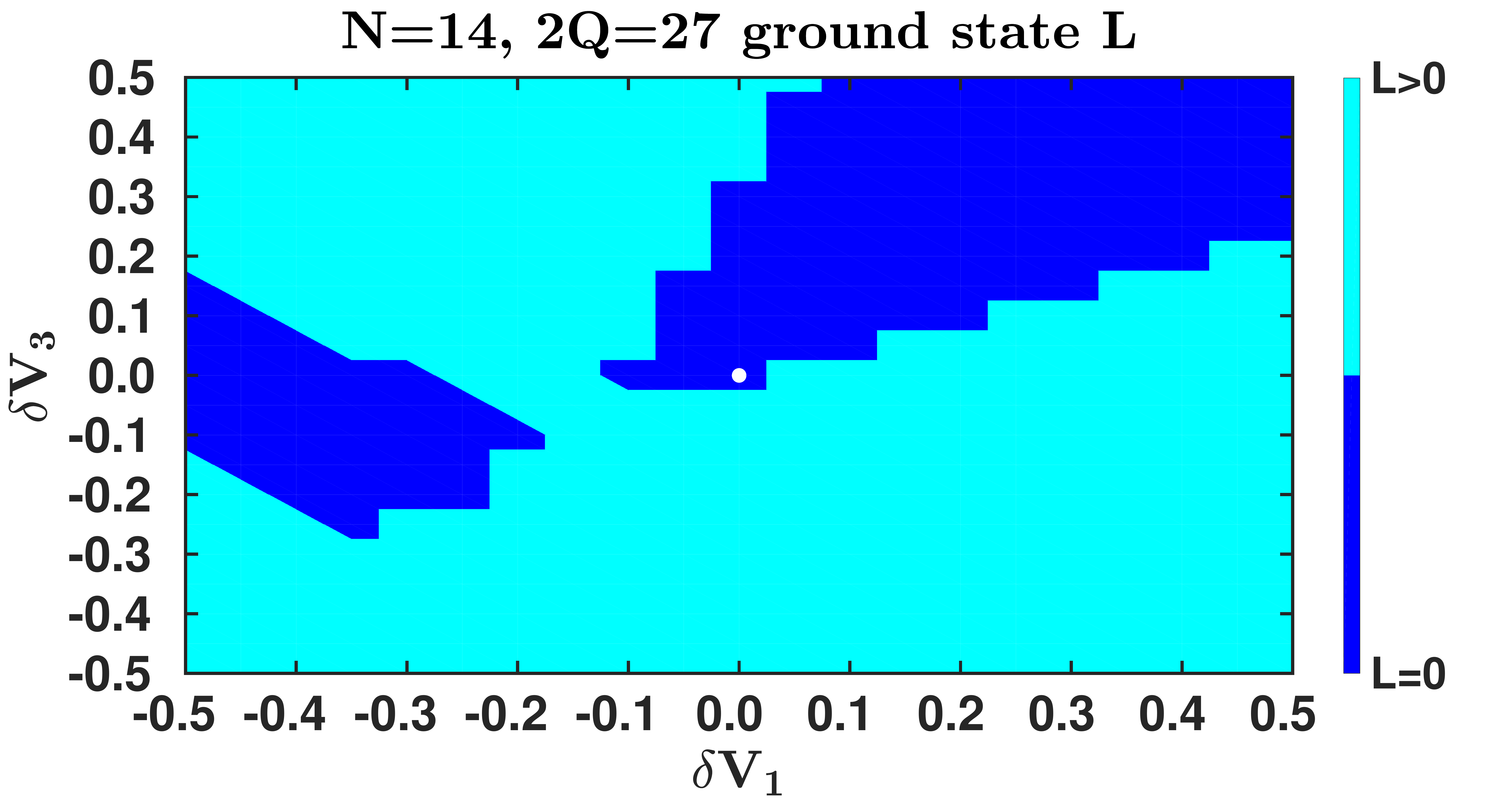} 
\includegraphics[width=0.3\textwidth]{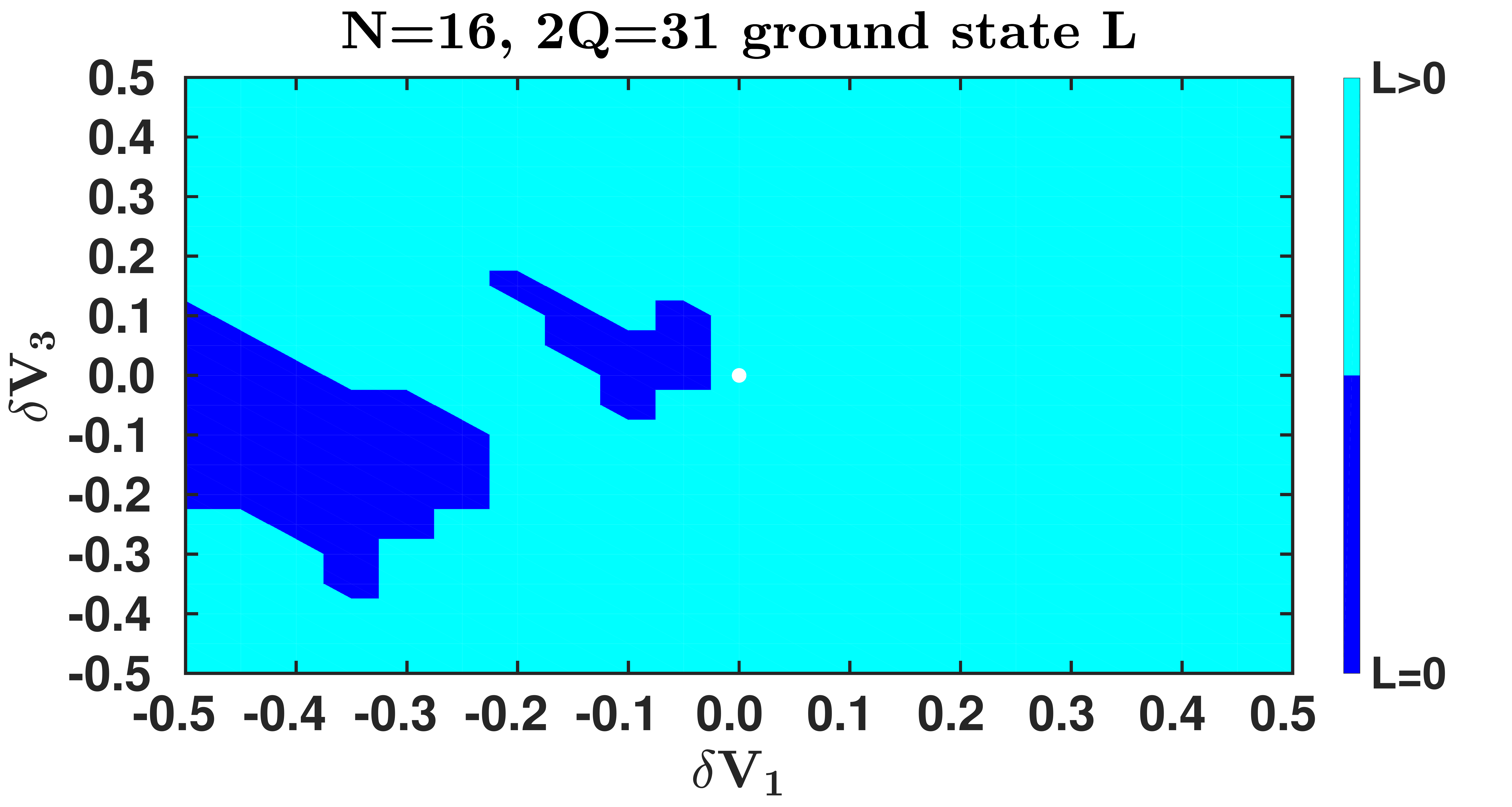} 
\caption{(color online) The ground state orbital angular momentum, $L$, obtained in the spherical geometry using exact diagonalization at the particle-hole symmetric Pfaffian flux for a model interaction (see Sec.~\ref{subsubsec:Coulomb_model} for a description of the interaction). Regions of parameter space where the ground state is uniform (i.e., has $L = 0$), and is therefore compatible with maintaining a finite energy gap in the thermodynamic limit, are indicated in dark blue shade. By perturbing $V_{1}$ and $V_{3}$ around the second Landau level Coulomb point (white dot in the center), we seek conditions where the PH-Pf state is stabilized. In the right panel of the top two rows we show the overlap of the particle-hole symmetric Pfaffian trial state $\mathcal{P}_{\rm LLL}[{\rm Pf}]^{*}J^{2}=\mathcal{P}_{\rm LLL}~\left[{\rm Pf}\left(\{(z_{i}-z_{j})^{-1}\}\right)\right]^{*}\prod_{i<j}(z_{i}-z_{j})^{2} $ with the ground state of the model interaction. In all cases, the PH-Pf does not appear to describe the ground state well.}
\label{fig:spectra_model_pps_phPf_perturb_around_SLL_Coulomb}
\end{center}
\end{figure*}

\begin{figure*}[ht]
\begin{center}
\includegraphics[width=0.3\textwidth]{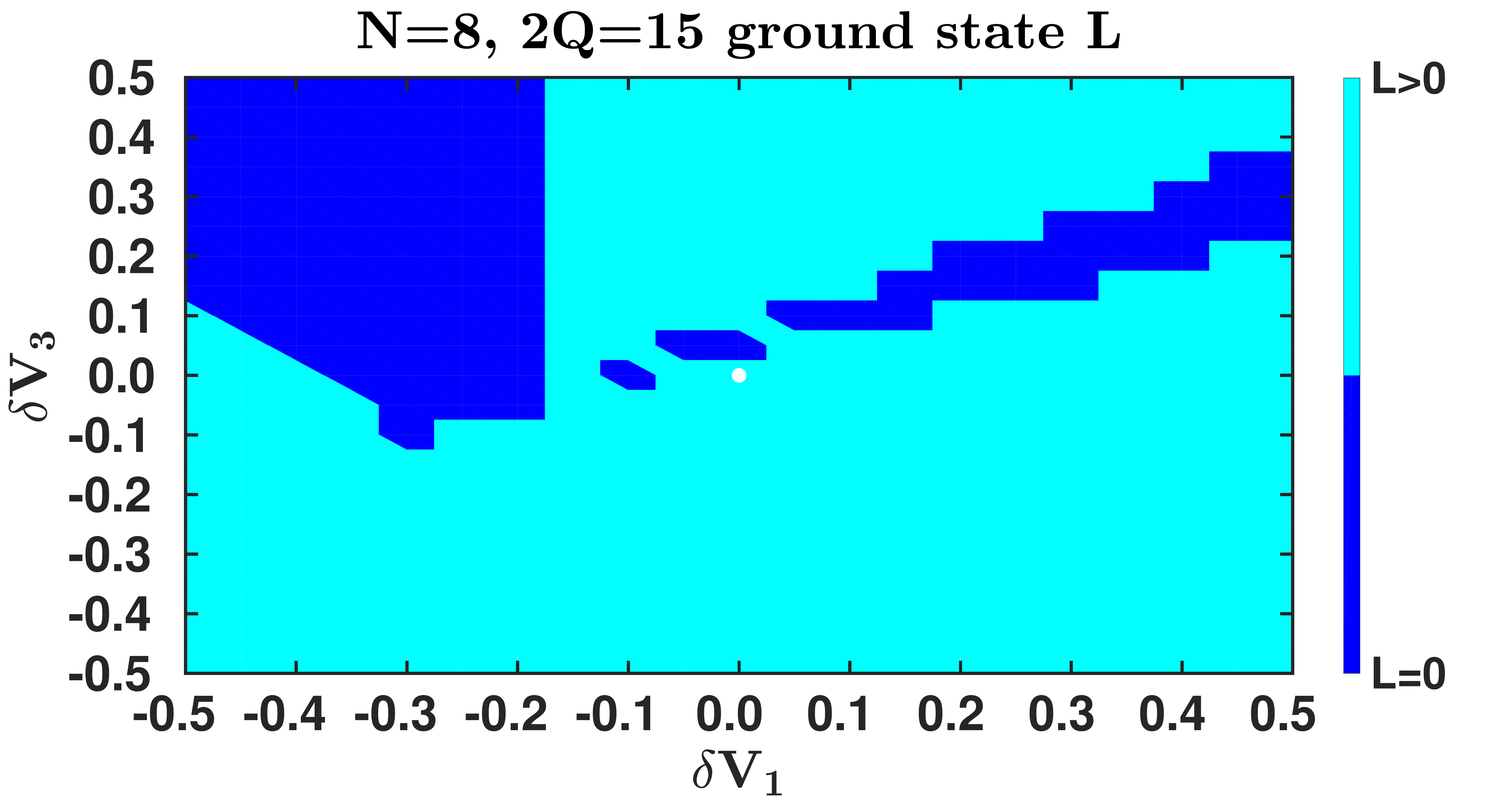} 
\includegraphics[width=0.3\textwidth]{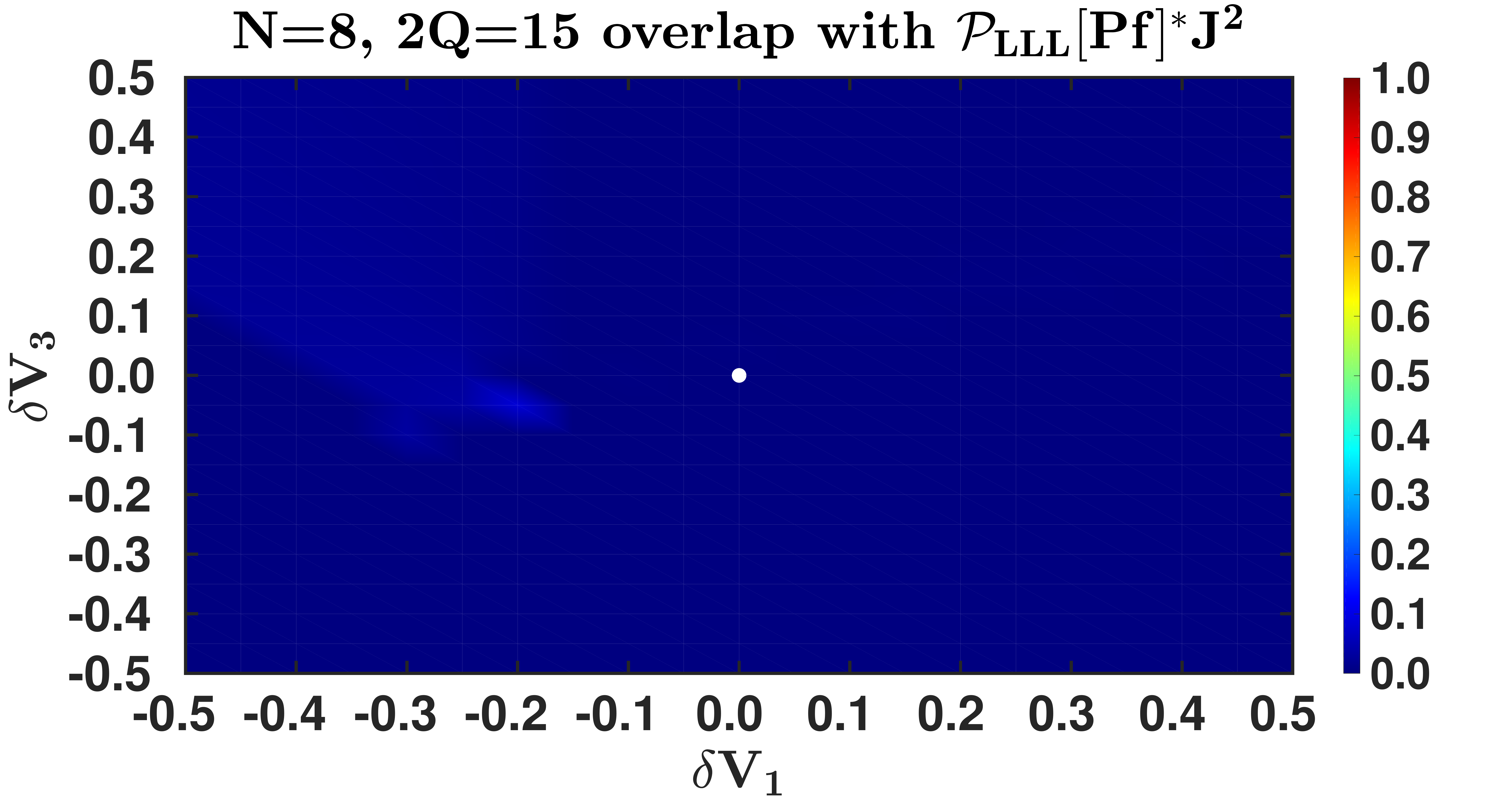} \\ 
\includegraphics[width=0.3\textwidth]{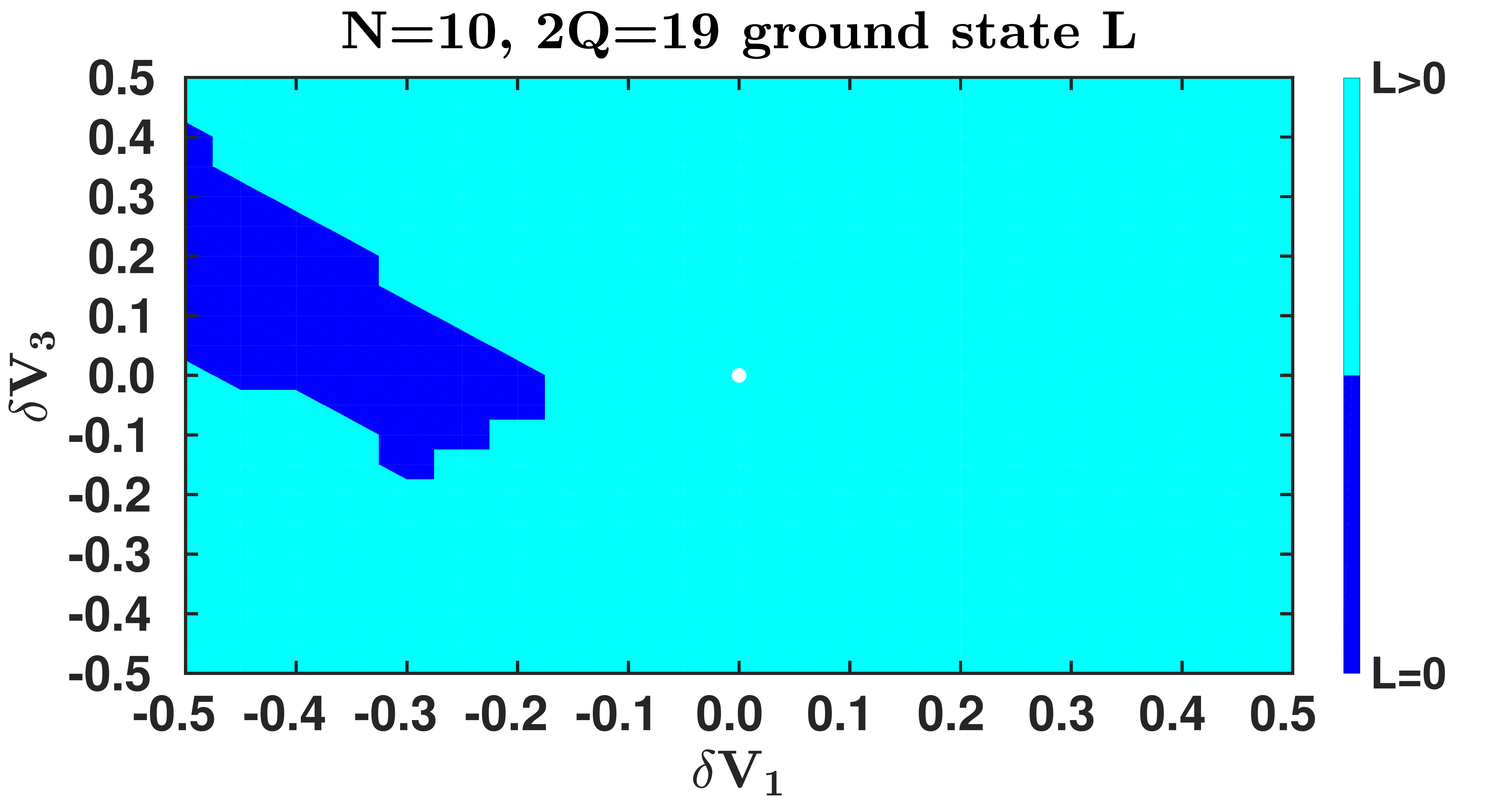} 
\includegraphics[width=0.3\textwidth]{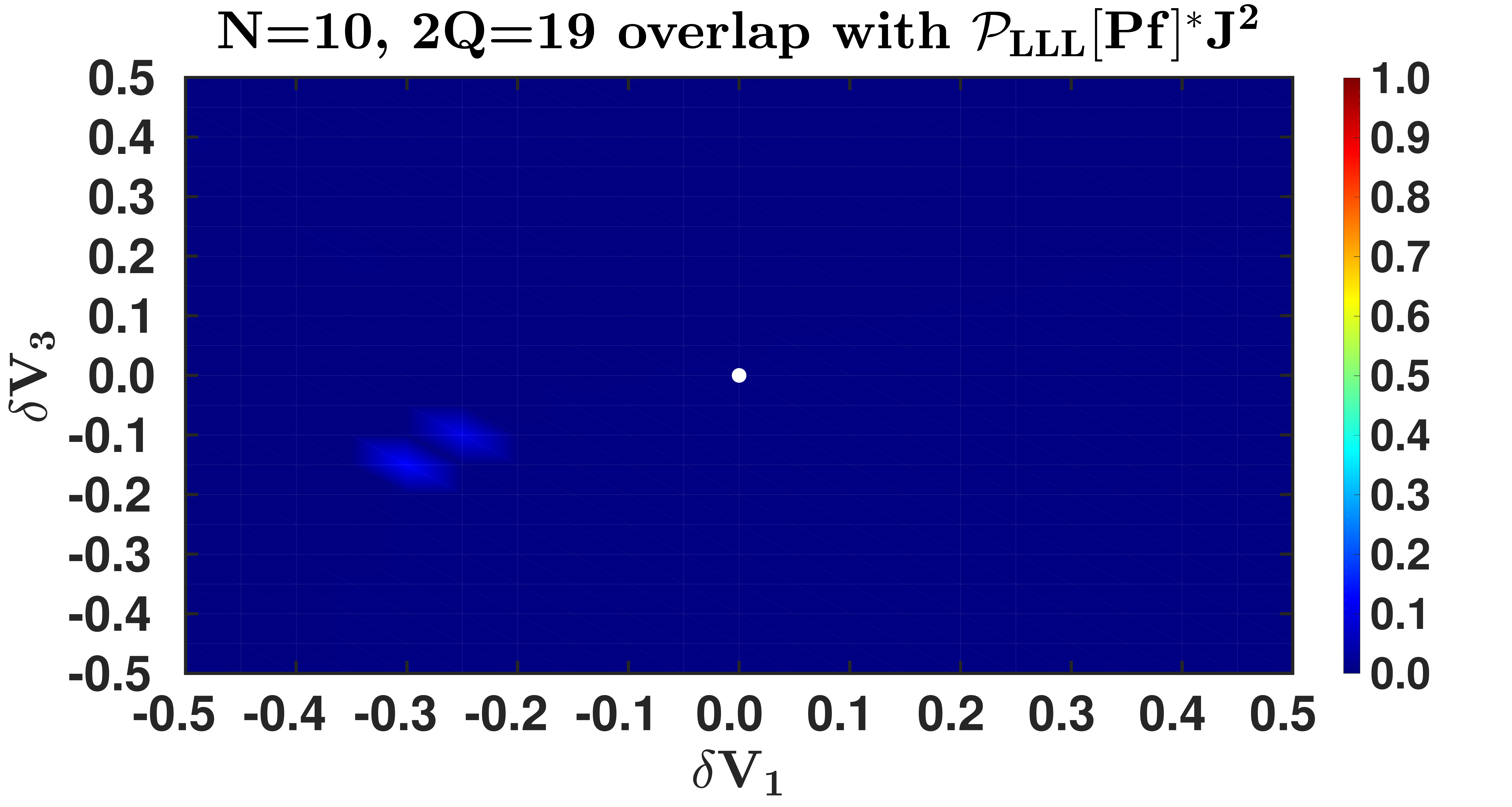} \\
\includegraphics[width=0.3\textwidth]{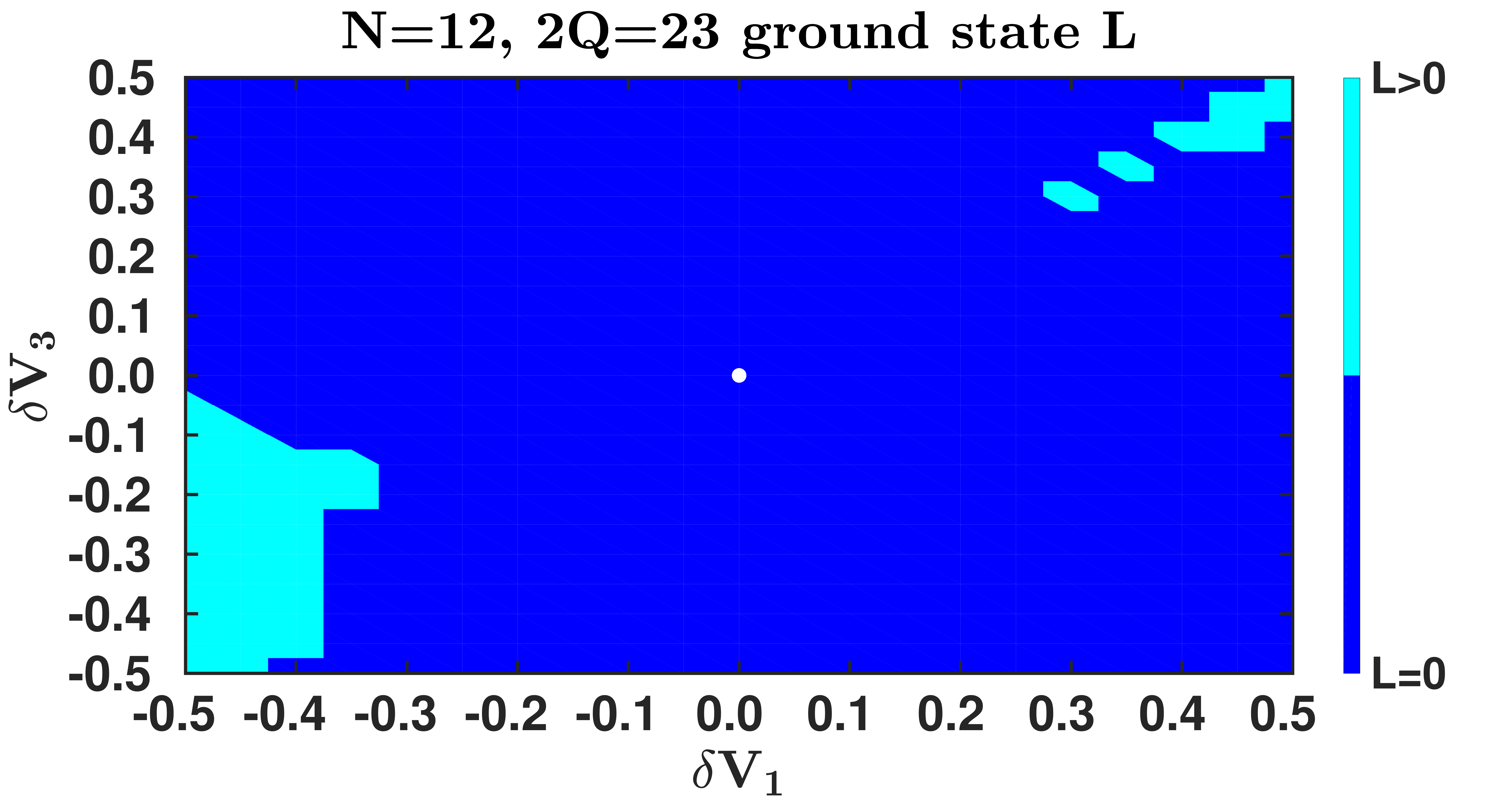}  
\caption{(color online) The ground state orbital angular momentum, $L$, obtained in the spherical geometry using exact diagonalization at the particle-hole symmetric Pfaffian flux for a model interaction (see Sec.~\ref{subsubsec:Coulomb_model} for a description of the interaction). Regions of parameter space where the ground state is uniform (i.e., has $L = 0$), and is therefore compatible with maintaining a finite energy gap in the thermodynamic limit, are indicated in dark blue shade. By perturbing $V_{1}$ and $V_{3}$ around the \emph{lowest} Landau level Coulomb point (white dot in the center), we seek conditions where the PH-Pf state is stabilized. In the right panel of the top two rows we show the overlap of the particle-hole symmetric Pfaffian trial state $\mathcal{P}_{\rm LLL}[{\rm Pf}]^{*}J^{2}=\mathcal{P}_{\rm LLL}~\left[{\rm Pf}\left(\{(z_{i}-z_{j})^{-1}\}\right)\right]^{*}\prod_{i<j}(z_{i}-z_{j})^{2} $ with the ground state of the model interaction. In all cases, the PH-Pf does not appear to describe the ground state well.}
\label{fig:spectra_model_pps_phPf_perturb_around_LLL_Coulomb}
\end{center}
\end{figure*}

In Fig.~\ref{fig:spectra_model_pps_phPf_perturb_around_SLL_Coulomb} (Fig.~\ref{fig:spectra_model_pps_phPf_perturb_around_LLL_Coulomb}) we show plots of the ground state total orbital angular momentum $L$, and the overlap with the candidate PH-Pf state at the particle-hole symmetric flux for different numbers of electrons for an interaction that is perturbed around the second (lowest) Landau level Coulomb point. Similar to our observations above, we do not find any region of parameter space where the ground state consistently occurs at $L = 0$ 
for all system sizes. Furthermore, in the region where the ground state is uniform for a particular $N$ (and hence could be compatible with maintaining an energy gap in the thermodynamic limit), we do not find a high overlap between the PH-Pf state and the ground state of the model interaction, consistently across all system sizes. These results suggest that even though the particle-hole symmetric Pfaffian trial state has a high overlap with its hole conjugate partner, it may not be realized as the ground state of a simple two-body interaction.

\section{Additional details for topological order of $\Phi^{2}_{2}$ }
\label{app:Phi2square}
Here we briefly provide some additional details for reading off the topological order of the state $\Phi^{2}_{2}$. As discussed in the main text, this state arises from a parton construction $b = f_4 f_5$, where $f_4$ and $f_5$ each form a $\nu = 2$ IQH state. This parton mean-field ansatz has an $SU(2)$ gauge symmetry, as any transformation 
$\left(f_4,f_5 \right)^{\rm T} \rightarrow W \left(f_4,f_5\right)^{\rm T}$, with $W \in SU(2)$, keeps $b$ invariant. The bulk effective theory can therefore be written in terms of an $SU(2)$ gauge field $A$, coupled to fermions $f_4$ and $f_5$. Integrating out $f_4$ and $f_5$ gives an $SU(2)_2$ Chern-Simons (CS) theory. The quasiparticles in this theory correspond to the particle / hole excitations in the mean-field IQH states, which are dressed by an $SU(2)_2$ CS gauge field. The flux attachment due to the CS term converts these excitations to anyons, whose non-Abelian fusion rules can be read off from the known properties of the $SU(2)_2$ CS theory. These are known to coincide with fusion rules of the Ising topological quantum field theory. 

Reading off the central charge is more delicate and is most clearly done by considering the boundary theory. The boundary theory for the mean-field state contains four chiral complex fermions, coming from the two $\nu = 2$ IQH states. These four fermions can be bosonized using non-Abelian bosonization, to give a $U(4)_1$ Wess-Zumino-Witten theory~\cite{DiFrancesco97}. However, not all of these degrees of freedom are physical; the physical degrees of freedom correspond to projecting out any fluctuations that are not invariant under the $SU(2)$ transformations. This can formally be performed using the coset construction~\cite{DiFrancesco97}. It follows that the boundary theory is a $U(4)_1 / SU(2)_2$ coset theory. The central charge of this theory is $c = c_{U(4)_1} - c_{SU(2)_2} = 4 - 3/2 = 5/2$. 

In summary, the state $\Phi^{2}_{2}$ describes a topological phase of bosons with central charge $c = 5/2$ and Ising fusion rules. The complex conjugate, $[\Phi^{2}_{2}]^*$, has reversed chirality and therefore describes a topological phase with central charge $c = -5/2$. 

\bibliography{biblio_fqhe}

\begin{thebibliography}{82}
\bibitem[{foo({\natexlab{a}})}]{footnote:reverse_flux}
\bibinfo{note}{For $n < 0$, note that the single parton forms an IQH state with
  filling factor $|n|$ in a reversed magnetic field, $-B_{|n|}$.}
\bibitem[{foo({\natexlab{b}})}]{footnote:two_thirds_ph_Laughlin}
\bibinfo{note}{The state $\Psi^{\rm CF}_{2/3}$ can also be thought of as the
  particle-hole conjugate of the Laughlin state at $\nu = 1/3$. For systems
  accessible to exact diagonalization we shall use the state $\mathcal{C}_{\rm
  ph}\Psi^{\rm L}_{1/3}$ for $\Psi^{\rm CF}_{2/3}$.}
\bibitem[{foo({\natexlab{c}})}]{footnote:Pfaffian}
\bibinfo{note}{Since the Pfaffian is non-zero only for even-dimensional
  matrices, the Moore-Read Pfaffian state is defined only for a system
  containing an even number of particles.}
\bibitem[{foo({\natexlab{d}})}]{footnote:Pfaffian_as_p_wave_paired_state}
\bibinfo{note}{Note the similarity with the Pfaffian state, where $\wp_{\rm Pf}
  = f_1 f_2 \psi_{l=1}$, where $\psi_{l=1}$ is in an angular momentum $l = 1$
  paired state.}
\bibitem[{foo({\natexlab{f}})}]{footnote:PHPf_no_P_LLL}
\bibinfo{note}{In the original proposal of Ref.~\cite{Zucker16} there is no
  $\mathcal{P}_{\rm LLL}$, as the authors of that study were interested in
  Landau level mixing.}
\end{thebibliography}
\bibliographystyle{apsrev_nourl}
\end{document}